\documentclass[%
reprint,
amsmath,amssymb,
aps,
]{revtex4-2}

\usepackage{units}
\usepackage{graphicx}
\usepackage{dcolumn}
\usepackage{bm}

\begin{document}
	
	
	\title{3D Fourier ghost imaging via semi-calibrated photometric stereo}
	
	\title{3D Fourier ghost imaging via semi-calibrated photometric stereo}
	
	\author{Ritz Ann Aguilar}
	\email{raguilar@nip.upd.edu.ph}
	\author{Nathaniel Hermosa}
	\author{Maricor N. Soriano}
	\affiliation{%
		National Institute of Physics, University of the Philippines Diliman, Quezon City 1101, Philippines\\
	}%
	
	
	
	
	\date{\today}
	
	\begin{abstract}
		We achieved three-dimensional (3D) computational ghost imaging with multiple photoresistors serving as single-pixel detectors using semi-calibrated lighting approach. 
		We performed imaging in the spatial frequency domain by having each photoresistor obtain the Fourier spectrum of the scene at a low spectral coverage ratio of 5\%. 
		To retrieve a depth map of a scene, we inverted, apodized, and applied semi-calibrated photometric stereo (SCPS) to the spectra. 
		At least 93.5\% accuracy was achieved for the 3D results of the apodized set of images applied with SCPS in comparison with the ground truth.
		Furthermore, intensity error map statistics obtained at least 97.0\% accuracy for the estimated surface normals using our method.
		Our system does not need special calibration objects or any additional optical components to perform accurate 3D imaging, making it easily adaptable. 
		Our method can be applied in current imaging systems where multiple detectors operating at any wavelength are used for two-dimensional (2D) imaging, such as imaging cosmological objects. 
		Employing the idea of changing light patterns to illuminate a target scene and having stored information about these patterns, the data retrieved by one detector will give the 2D information while the multiple-detector system can be used to get a 3D profile.
		
		\vspace{1em}
		\noindent\textbf{© 2021 Optica Publishing Group.} 
		One print or electronic copy may be made for personal use only. Systematic reproduction and distribution, duplication of any material in this paper for a fee or for commercial purposes, or modifications of the content of this paper are prohibited.
		
	\end{abstract}
	
	\maketitle
	

	\section{Introduction}
	\label{sec:intro}
	Single-pixel cameras are gaining prevalence in the field of imaging because they break the limits set by a conventional camera, particularly in the non-visible spectrum. 
	This is made possible through the concept of ghost imaging (GI), an unconventional imaging technique which initially relied on light correlations from data obtained by two light detectors \cite{Pittman1995}.  
	Although it originated from quantum sensing with initial experiments using entangled photons, the field has later on expanded using classical-state light source such as pseudo-thermal light \cite{Bennink2002,Valencia2005,Ferri2005,Zhang2014,Devaux2017} and broadband sources \cite{Bennink2004,Gatti2004}. Advancements in the technique introduced computational ghost imaging (CGI) that uses a single non-pixelated light detector \cite{Shapiro2008} to obtain two-dimensional images of a scene using compressed sensing techniques \cite{Candes2006, Donoho2006, Duarte2008a}.  
	A variety of applications then emerged that include non-visible light imaging applied to different fields \cite{Edgar2015_1,Stantchev2016,Gibson2017,Kim2020,Kingston2018,Zhang2018a} with additional experimental demonstrations on ghost spectroscopy \cite{Scarcelli2003, Janassek2018, Olivieri2018}.
	
	In 2009, Bromberg et al. first introduced the idea of three-dimensional (3D) imaging applications for GI \cite{Bromberg2009} citing its depth-resolving and diffraction imaging capability. 
	Clemente et al. later on presented the concept of digital ghost holography \cite{Clemente2012}.
	Ghost tomography using hard x-rays of optically opaque 3D objects has also been demonstrated with a ghost imaging setup using a classical-state source \cite{Kingston2018}.
	An interesting application on 3D CGI exploited time-of-flight (TOF) approach for remote sensing and atmospheric imaging \cite{Howland2011,Howland2013,Chen2013,Yu2015,Gong2016}.
	In 2016, Sun et al. used a modified TOF approach based on short-pulsed structured illumination combined with compressive sensing for low-cost real-time 3D imaging \cite{Sun2016}.
	Another recent study on non-line-of-sight 3D imaging also used TOF coupled with back-projection imaging algorithms to achieve real-time imaging \cite{Musarra2019}.
	
	Here, we offer a simple, cheap, and adaptable approach to 3D ghost imaging by extending our previous work on low-cost ghost imaging (see Ref.~\cite{Aguilar2019}) using photoresistors as single-pixel detectors (SPDs). 
	We show that shading information is preserved in the Fourier domain and our low-cost ghost imaging system is capable of 3D imaging using a semi-calibrated photometric stereo (SCPS) algorithm. 
	This differs from existing 3D CGI systems that use stereo vision or shape-from-shading technique done in Ref.~\cite{Sun2013} or described in Ref.~\cite{Sun2019} because far fewer measurements for each detector are needed since imaging is performed in the Fourier domain.
	As a proof of concept, we apply the method to test objects without sharp edges. 
	This allows us to perform 3D imaging at a low spectral coverage ratio, $\alpha$, of 5\% because of the absence of complex or high spatial frequency features.
	Additionally, these photometric stereo setups usually follow the assumption of a fully-calibrated system with sensors receiving the same amount of light or if not, a mirror or chrome sphere for estimating light parameters is used.
	We show here that such special calibration objects are not needed with the SCPS algorithm.
	
	Moreover, our 3D imaging system does not need additional optical components such as lenses or gratings unlike the techniques used in Refs.~\cite{Zhang2016, Zhang2018, Teng2020} which also image in the Fourier domain.
	The setups in Refs.~\cite{Zhang2016} and \cite{Zhang2018} mimic a fringe projection profilometry system which employs the usual phase-to-height calibration using objects with known dimensions.
	Teng et al. \cite{Teng2020} even focus more on high-speed imaging by using a mode-lock laser for the light source and a time-encoded projector to generate illumination patterns, with a calibration part that involves imaging a standard flat and spherical surface.
	A recent study on single-pixel 3D imaging also did not employ additional hardware but made use of multiplexed illumination and a convolutional neural network to decode both reflectance and depth information \cite{Wang2021}. 
	
	Other recent studies on 3D CGI involving photometric stereo make use of Hadamard patterns of low and high spatial resolutions that are based on the surface normal variations of the object as was done by Qian et. al \cite{Qian2019}.
	The same basis set was used by Zhang et al. in Ref.~\cite{Zhang2016a} coupled with a spatial light modulator of high modulation rate for real-time imaging. 
	Soltanlou et al. image a 3D object through scattering media in Ref.~\cite{Soltanlou2019} while Zhang et al. improve noise immunity of 3D CGI systems using sub-pixel displacement in Ref.~\cite{Zhang2019}.
	As for other basis sets, Haar wavelets are used as illumination patterns by Xi et al. combined with a bi-frequency projecting system for high-contrast and high-resolution 3D ghost imaging \cite{Xi2019}.
	
	In this paper, we present a three-step process to obtain 3D reconstructions of objects. 
	First, we use the low-cost ghost imaging technique performed in the Fourier domain introduced in Ref.~\cite{Aguilar2019} to obtain several 2D images of a scene using multiple photoresistors. 
	Second, we improve on the quality of the images captured by the photoresistors by applying an apodization technique on the retrieved Fourier spectra.
	Third, we apply a semi-calibrated photometric stereo approach that does not need a special calibration object to estimate the lighting strength and direction for each detector. 
	This also addresses the problem of differing sensitivities (although minimal) of the photoresistors used in the experiment. 
	We believe this is the first time the semi-calibrated photometric stereo combined with apodization technique is used for 3D reconstruction in computational ghost imaging.  
	
	\section{Semi-calibrated photometric stereo}
	\label{sec:scps}
	Photometric stereo has been demonstrated as a powerful technique in retrieving depth information using the concept of shape-from-shadow \cite{Woodham1980, Horn1989}.  
	Several images with different illuminations by different light sources are used to reconstruct the 3D image of a scene. 
	These images are captured at a fixed viewpoint by a digital camera. 
	Invoking Helmholtz reciprocity, the sensor and the light source may be interchanged. 
	This inverse principle is the core idea of 3D CGI via photometric stereo that uses several SPDs and a single light source.
	
	\begin{figure}[btp]
		\centering
		\includegraphics[width=\linewidth]{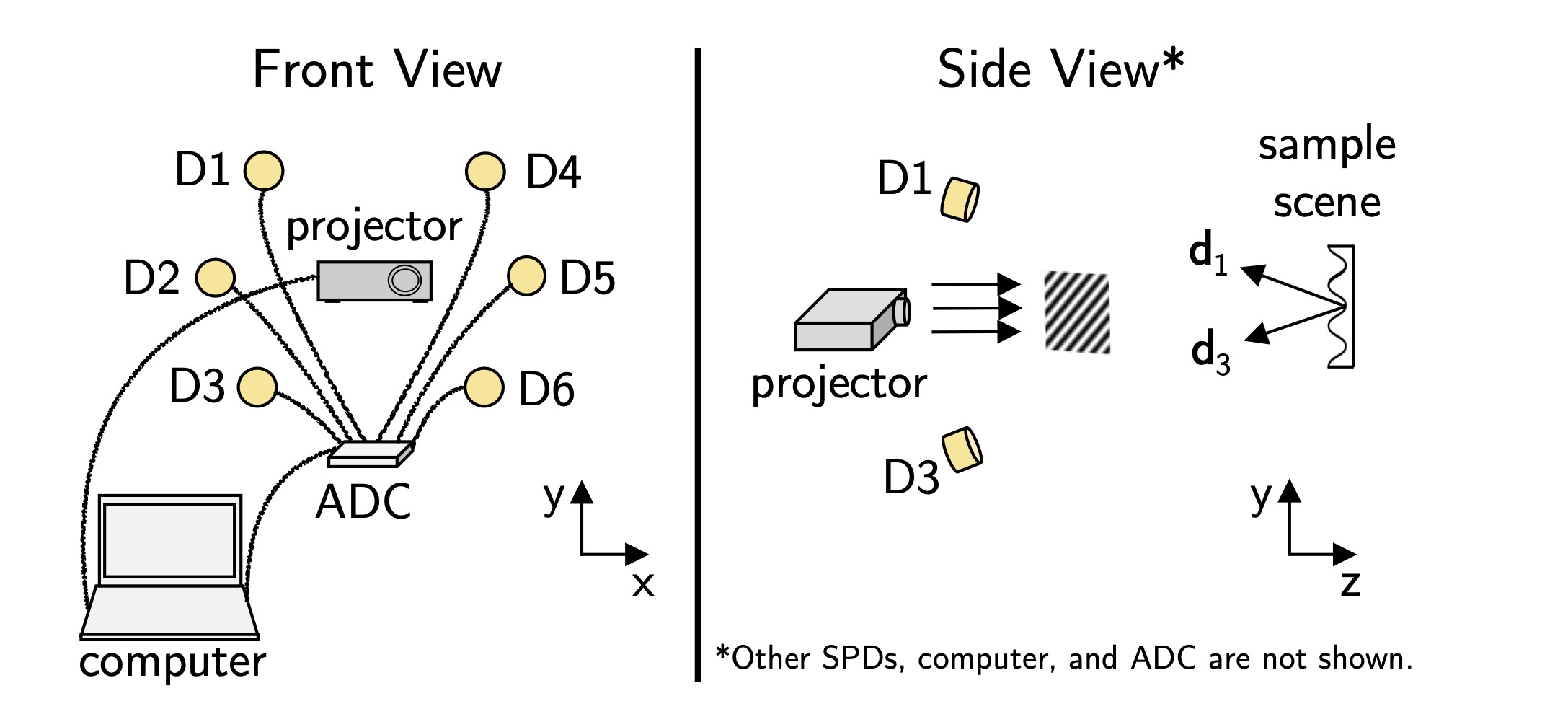}
		\caption{
			3D Fourier ghost imaging experiment setup. Single-pixel detectors D1 to D6 (photoresistors) are strategically placed in the positions shown to ensure enough shading information in the resulting images. A commercial projector is placed farther back which illuminates the scene with sinusoidal fringe patterns of varying frequencies. The projection and capturing are synchronized using a computer while the signal from the detectors are digitized using an analog-to-digital converter.}
		\label{fig:setup}	
	\end{figure}
	
	In a CGI photometric stereo setup wherein there are multiple detectors (Fig.~\ref{fig:setup}), each detector $i$ measures different intensities $I_{i,j}$ where $j$ represents a single pixel in the $i^{\text{\text{th}}}$ image. We want to recover the scene properties corresponding to the albedo or surface reflectivity $\beta_j$ and the surface normal $\textbf{n}_j$. Letting $\mathbf{g}_j = \beta_j \mathbf{n}_j$ which we call the albedo-scaled surface normal, the image recovered from the $i^{\text{th}}$ detector is
	\begin{equation}
	I_{i,j} = \mathbf{d}_i^\top \mathbf{g}_j,
	\label{eq:psintensity}
	\end{equation}
	where $\textbf{d}_i$ corresponds to the detector position vector.
	We additionally define $\textbf{d}_i$ as
	\begin{equation}
	\mathbf{d}_i = \gamma_i \hat{d}_i,
	\label{eq:lightstrength}
	\end{equation}
	where $\gamma_i$ represents the lighting strength on the $i^{\text{th}}$ detector and $\hat{d}_i$ represents the unit vector corresponding to the detector's location.
	
	Considering all detectors, Eq.~(\ref{eq:psintensity}) may be recast into matrix form
	\[
	\begin{bmatrix}
	I_1 \\
	\vdots \\
	I_j
	\end{bmatrix}
	= 
	\begin{bmatrix}
	\mathbf d_1 \\
	\vdots \\
	\mathbf d_i
	\end{bmatrix}
	\mathbf g_j.
	\]
	that we can consequently write as
	\begin{equation}
	\mathbf{I}_j = \mathbf{D}^\top  \mathbf g_j.
	\label{eq:psintensitymatrix}
	\end{equation}
	Here, $\mathbf{I}_j \in \mathbb{R}^{M \times 1}$ is a matrix concatenation of all $I_{i,j}$ (with $i \in [1,M]$ for $M$ images) representing all measured intensities, $\mathbf{D}^\top \in \mathbb{R}^{M \times 3}$ are either fully or partially calibrated lighting parameters from all detectors, and $\textbf{g}_j$ is the unknown scene property we want to retrieve.
	
	Solving for $\mathbf{g}_j$ using a least-squares solution approach, we have
	\begin{equation}
	\mathbf g_j = (\mathbf D \mathbf{D}^\top)^{-1} \mathbf D \mathbf I_j,
	\label{eq:lsq}
	\end{equation}
	from which the	 surface normal may be computed as
	\begin{equation}
	\mathbf n_j = \dfrac{\mathbf g_j}{|| \mathbf g_j ||},
	\label{eq:normal}
	\end{equation}
	with $|| \mathbf g_j || = \beta_j$.
	
	In a partially calibrated lighting environment such as in our case where the lighting strength $\gamma_i$ in Eq.~(\ref{eq:lightstrength}) is not known but the unit vector $\hat{d}_i$ is known, we estimate the unknown parameter from the images themselves. 
	Cho et al. \cite{Cho2012} called this the semi-calibrated photometric stereo.
	It is fit for scenarios where there is (i) varying light intensity conditions and exposures, and/or (ii) limited control over sensors.
	It was primarily intended for conventional photometric stereo setups having multiple light sources and one sensor, however, it can also be used for 3D CGI setups.
	Given that the detectors have different characteristics and that we cannot entirely control the amount of light received by each, SCPS is applicable to our setup.
	
	To solve for the unknown, we invoke a minimization technique of a  cost function involving the lighting parameters \cite{Xiong2015},
	\begin{equation}
	\text{cost}(\{\gamma_i \}) = 
	\begin{array}{ll} \sum\limits_j \min\limits_{\mathbf{g}_j} \| \mathbf{I}_j - \mathbf{D(\Gamma)}^\top \mathbf{g}_j \|^2, \end{array}
	\label{eq:cost1}
	\end{equation}
	where $\mathbf{D(\Gamma)} = \left[\gamma_1 \hat{d}_1 \quad \cdots \quad \gamma_n \hat{d}_M \right]$ with $\mathbf{\Gamma} \in \mathbb{R}^{1 \times M}$, a one-dimensional vector containing the optimal light intensity measured by each detector. 
	
	The function $\mathbf{g}_j$ is obtained in the same manner as Eq.~(\ref{eq:lsq}), reducing the cost function to
	\begin{equation}
	\text{cost}(\{\gamma_i \}) = 
	\begin{array}{ll} \sum\limits_j \| \mathbf{I}_j - \mathbf{D(\Gamma)}^\top 
	(\mathbf{D(\Gamma)} \mathbf{D(\Gamma)}^\top)^{-1} \mathbf {D(\Gamma)} \mathbf I_j \|^2, \end{array}
	\label{eq:cost2}
	\end{equation}
	that can be solved using a non-linear least squares approach such as the Levenberg-Marquardt algorithm \cite{Levenberg1944, Marquardt1963,Kanzow2004}. 
	This gives us the optimal value of $\mathbf{\Gamma}$ for the depth map estimation.
	
	We then employ the well-known Frankot-Chellappa algorithm \cite{Frankot1988} to solve for the surface elevation
	\begin{equation}
	z = f(x,y),
	\label{eq:elevation}
	\end{equation}
	where $(x,y,z)$ are the object coordinates.
	Letting $(p,q)$ be the gradient of the above function, we may define the surface normal as $[p,q,-1]$.
	Explicitly, these parameters are related to the coordinates of the normal vector
	\begin{equation}
	z_x = \frac{\partial f(x,y)}{\partial x} = -\frac{n_{x}}{n_{z}} = p, \qquad
	z_y = \frac{\partial f(x,y)}{\partial y} = -\frac{n_{y}}{n_{z}} = q,
	\end{equation}
	where $(n_x, n_y, n_z)$ were retrieved using Eq.~(\ref{eq:normal}).
	Simultaneously, the distance error is minimized according to
	\begin{equation}
	\iint|z_x-p|^{2}+|z_y-q|^{2} dx dy.
	\label{eq:disterror}
	\end{equation}
	The surface geometry is obtained by letting Eq.~(\ref{eq:elevation}) be a linear combination of discrete Fourier basis functions $\phi(x, y, \boldsymbol{\omega})$ to ensure integrability,
	\begin{equation}
	f(x, y)=\sum_{(\boldsymbol\omega} {C}(\boldsymbol{\omega}) \phi(x, y, \boldsymbol{\omega})
	\end{equation}
	where $\boldsymbol{\omega} = (\omega_x,\omega_y)$ is a 2D index in the Fourier plane.
	The optimal coefficients that minimize Eq.~(\ref{eq:disterror}) are
	\begin{equation}
	{C}(\omega)=\frac{-j \omega_{x} C_{x}(\omega)-j \omega_{y} C_{y}(\omega)}{\omega_{x}^{2}+\omega_{y}^{2}}
	\label{eq:Comega}
	\end{equation}
	with $C_x(\omega)$ and $C_y(\omega)$ being Discrete Fourier Transforms (DFT) of $p$ and $q$, i.e., $C_x(\omega) = \mathcal{F}\left\{p\right\}$ and $C_y = \mathcal{F}\left\{q\right\}$.
	Finally, the depth map is estimated by taking the inverse DFT of Eq.~(\ref{eq:Comega}).

	
	\section{Experiment}
	\label{sec:experiment}
	Six photoresistors (\textit{GL12528, 12 mm $\times$ 1 mm sensitive area}), labeled D1 to D6, were strategically placed at an approximate distance of $\unit[30]{cm}$ from the scene to ensure enough shading information in the resulting images as shown in Fig.~\ref{fig:setup}.
	With the lensless imaging setup, the SPDs must be relatively near the scene to get a good signal-to-noise ratio.
	A commercial projector (\textit{Acer K11 DLP 200}), placed  $\unit[1]{m}$ behind the detectors, is used to illuminate the scene with sinusoidal light patterns while the photoresistors measure the back-scattered light.
	A computer (\textit{16-GB RAM, Intel Core i7-4710MQ}) generates the patterns and consequently performs 3D reconstructions while the signal from each photoresistor is digitized using an \textit{Arduino Uno Rev3} microcontroller.
	
	\begin{figure}[t]
		\centering
		\includegraphics[width=0.85\linewidth]{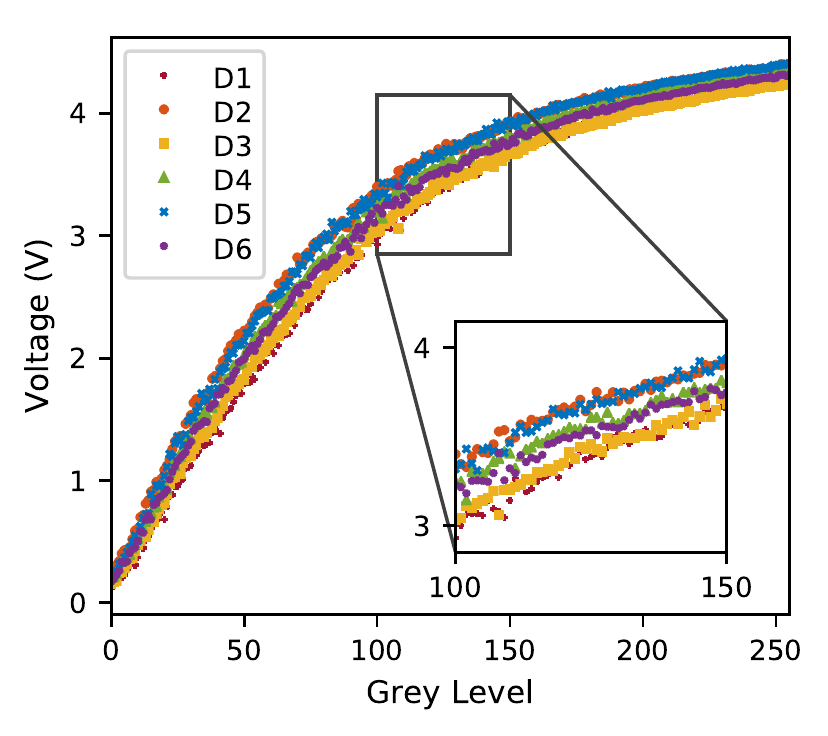}
		\caption{The response curve (voltage as a function of projector grey level) was plotted for all detectors. The inset on the lower right shows a close-up response curve of the detectors in the chosen region. Labels D1 to D6 correspond to each detector as described in Fig.~\ref{fig:setup}. }.
		\label{fig:calib}	
	\end{figure}
	
	In our experiment, each detector first performs imaging in the Fourier domain. The sets of phase-shifted sinusoidal fringe patterns, with  spatial frequencies covering $\alpha = 0\%-5\%$ in the Fourier domain, are projected onto the scene to obtain the coefficients of the spectrum \cite{Zhang2015}. 
	Each set has 3 sinusoidal patterns of the same spatial frequency and with $0$, $2\pi/3$, and $4\pi/3$ phase shifts.
	The inverse fast Fourier transform algorithm is then applied to each Fourier spectrum to obtain the actual image. From the combined images retrieved by all detectors, we then apply the SCPS approach to obtain the 3D image of the scene.
	
	\begin{figure}[tbp]
		\centering
		\includegraphics[width=0.8\linewidth]{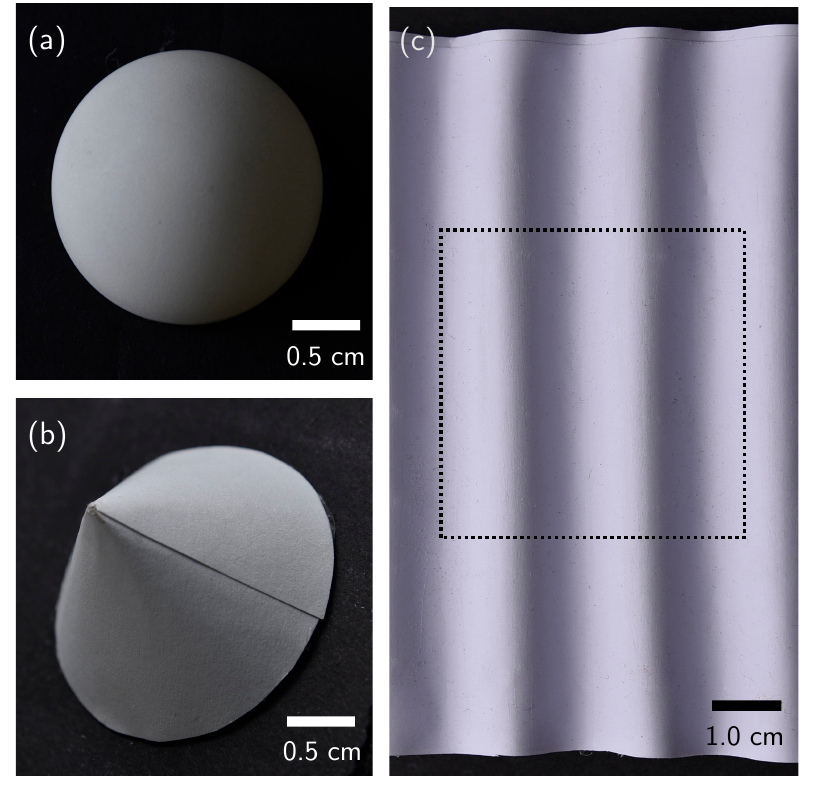}
		\caption{Photographs of the objects used for the 3D ghost imaging experiment: a hemisphere (a), a cone (b), and a 3D sine wave (c). Each object has a matte surface to minimize specular reflection. The boxed area in (c) shows the region of interest for the 3D reconstruction of the sine wave object. The top view of the cone was imaged by the detectors for the 3D reconstruction.}
		\label{fig:objects}	
	\end{figure}
	
	We characterized all photoresistors and synchronized the fringe projection and signal recording. 
	Figure~\ref{fig:calib} shows the response curves obtained from the six detectors. 
	For each detector, voltage was plotted as a function of increasing grey level which is proportional to light intensity, projected on a white plane placed at the location of the scene. 
	As expected, the photoresistors have a nonlinear response to increasing light intensity. 
	In general, we saw minimal difference in their response curves but when inspected closely (inset on lower right of Fig.~\ref{fig:calib}), not all of these detectors have overlapping data points. 
	This of course has an effect on the magnitude of the light they will receive at their location and in turn affects the amount of shading information in the resulting images. 
	\textit{To compensate for this, SCPS is used -- it does not only take into account the detector positions but also the light strength they are receiving.} 
	We reiterate that in most fully-calibrated photometric stereo systems, a mirror or chrome sphere for estimating light strength is utilized but with the SCPS algorithm, such special calibration objects are unnecessary.

	\begin{figure*}[tbp]
		\centering
		\includegraphics[width=0.9\linewidth]{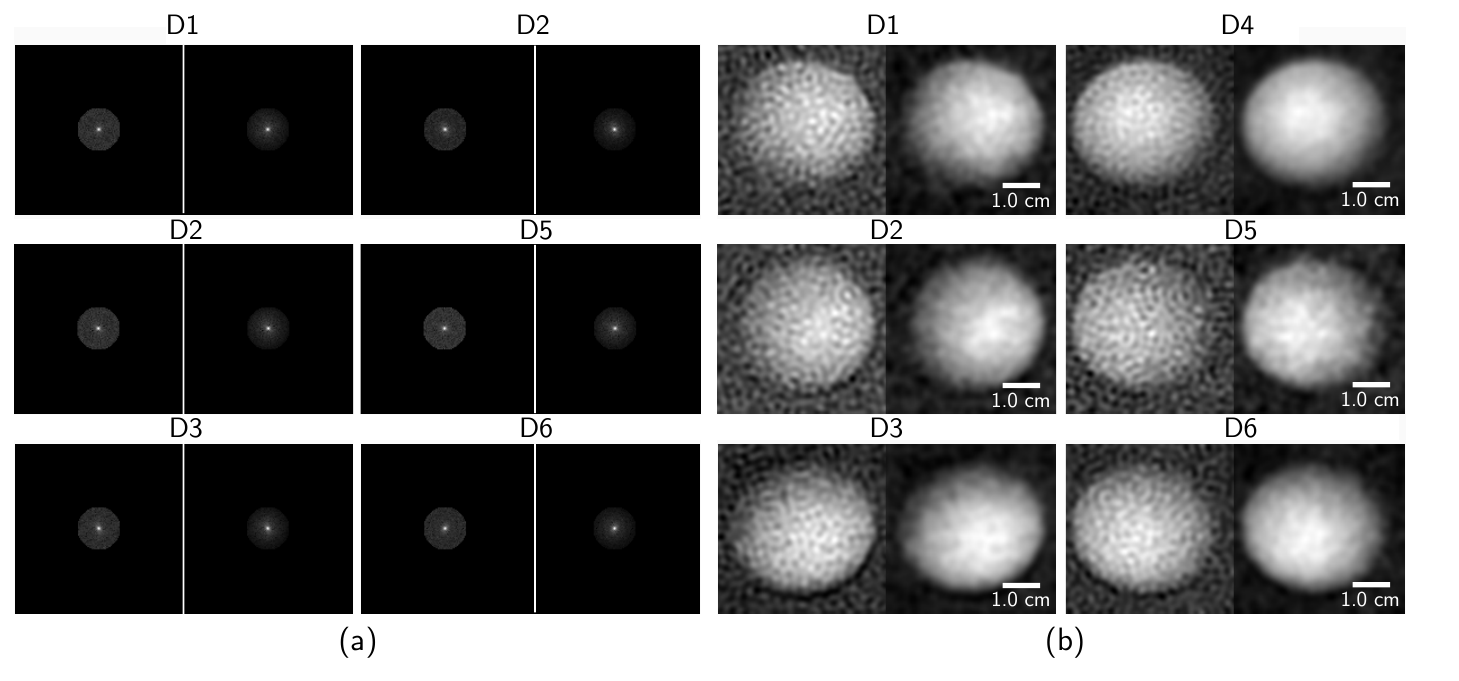}
		\caption{Fourier spectra (a) at 5\% spectral coverage ratio and corresponding images upon Fourier inversion (b) as captured by the six single-pixel detectors for the hemisphere object. Each left-and-right pair corresponds to before-and-after apodizing the spectrum and the labels D1 to D6 to each detector in Fig. \ref{fig:setup}. The images have a size of $\unit[150]{pixels} \times \unit[150]{pixels}$.}
		\label{fig:spherescpectra}	
	\end{figure*}
	
	\begin{figure}[tbp]
		\centering
		\includegraphics[width=\linewidth]{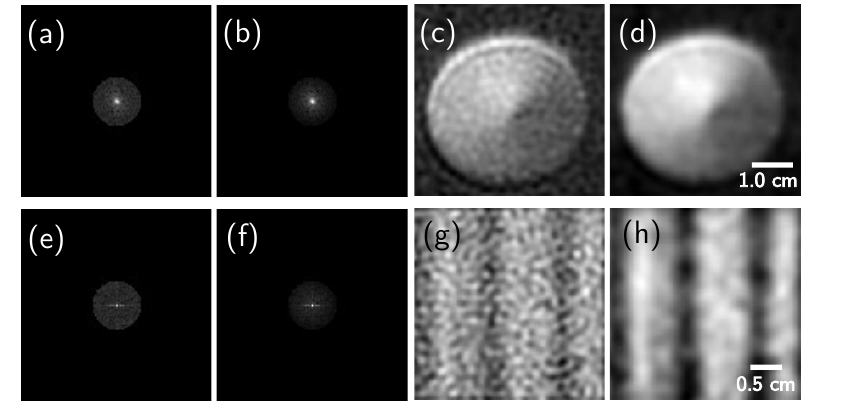}
		\caption{Sample retrieved Fourier spectra of one of the detectors before (a, e) and after (b, f) apodization for the cone and sine wave object, respectively. The corresponding images for (a, b) upon inversion are shown in (c, d) while those of (e, f) are shown in (g, h).}
		\label{fig:apodized}	
	\end{figure}
	
	As a proof of concept of 3D Fourier ghost imaging (FGI) using SCPS, we only used test objects with simple 3D structure during the experiment which include a hemisphere, a cone, and a  3D sine wave (Fig.~\ref{fig:objects}). 
	The hemispherical object and the cone have an estimated radius of $\unit[2.0]{cm}$ while the cone has a height of $\unit[2.1]{cm}$. 
	The amplitude and wavelength of the sine wave are $\unit[0.5]{cm}$ and $\unit[2.0]{cm}$, respectively. 
	All of the objects have matte surfaces to reduce specular reflectance and assure a Lambertian surface. 
	Intuitively, the placement and orientation of the detectors relative to the objects affect the measured back-scattered light. 
	This also depends on the topographical texture of the object. 
	For example, orienting the waves horizontally as shown in the side view of the setup in Fig.~\ref{fig:setup} will result in less shadows in the images produced by detectors D2 and D5. Consequently, orienting the waves in a diagonal manner will result to little shadow information from either D1 and D6 or D3 and D4 depending on the wavelength alignment.
	The sine wave was oriented vertically (Fig.~\ref{fig:objects}c) such that we have enough shadow information from all detectors.
	For the hemisphere and cone, we took their top view images.
	
	\section{Results and Discussion}
	
	Each detector gives a different Fourier spectrum by virtue of its position.
	For the reconstruction, we set $\alpha$ to $5\%$ of the entire Fourier spectrum and the size of the images to $150 \times \unit[150]{pixels}$ covering roughly $4.3 \times \unit[4.3 \pm 0.1]{cm^2}$ of the actual scene.
	The image reconstruction algorithm utilizes the idea that natural
	images are sparse in the spatial frequency domain \cite{Taubman2002}.
	Statistics show that their most informative spatial frequency bands are contained within the $5\%$ spectral coverage, or $10\%$ for edge retrieval \cite{Bian2016}.
	Considering 3 step phase-shifting illumination and symmetry of the Fourier spectrum, a $5\%$ spectral coverage for the set image size is equivalent to 1688 single-pixel measurements.
	Accounting for delay time between fringe projection and signal acquisition (set to $\unit[25]{ms}$ with the photoresistor's response time as the limiting factor), bit fluctuations of the analog-to-digital conversion (we averaged over 20 trials), and sequential reading of the SPDs, it took ${\sim}1.7$ hours to produce each set of six images for a single object or ${\sim}5$ minutes for one trial.
	
	\begin{figure}[tbp]
		\centering
		\includegraphics[width=0.9\linewidth]{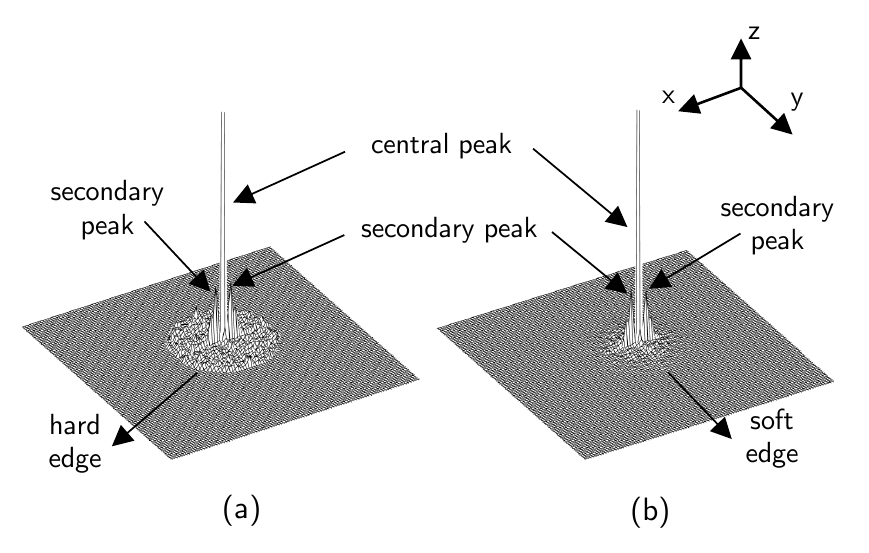}
		\caption{Fourier spectrum in 3D of the sine wave object before (a) and after (b) apodization. The secondary peaks, representing the frequency of the sine wave, beside the central peak (DC term) were preserved while the hard edge was softened out. We limit the $z$-axis, cutting off the central peak, to emphasize the secondary peaks and other details in both spectra.}
		\label{fig:sine3Dspectra}	
	\end{figure}
	
	We improve on the quality of an image obtained by a photoresistor by apodizing the retrieved Fourier spectrum, an optical filtering technique that smoothens out the edges of an aperture \cite{Slepian1965, McCutchen1969, Thomas2001}. 
	In particular, since we used a circular sampling approach, we multiply a centered Gaussian filter to each spectrum of the form 
	\begin{equation}
	h(\xi,\eta) = \dfrac{1}{2\pi\sigma^2} \exp\left(-\dfrac{(\xi/K)^2 + (\eta/L)^2}{2\sigma^2} \right),
	\end{equation}
	where $\xi \in [-K/2,K/2]$ and $\eta \in [-L/2,L/2]$ with $K$ and $L$ being the number of pixels along the $\xi$ and $\eta$ axes of the image, and $\sigma$ is the standard deviation of the Gaussian distribution.
	Note that one must be careful in choosing the extent of apodization which is equivalent to setting the radius of the Gaussian filter as dictated by $\sigma$.
	\url{Visualization 1} shows how the Gaussian filter looks like as we vary $\sigma$ and its effect on one spectrum and corresponding reconstructed image.
	Since we only want to soften the edges of the mask while ensuring important features of the spectra are preserved, we set the value of $\sigma$ as equal to the spectral coverage ratio used, $\sigma = \alpha = 0.05$.
	
	\begin{figure*}[t]
		\centering
		\includegraphics[width=0.9\linewidth]{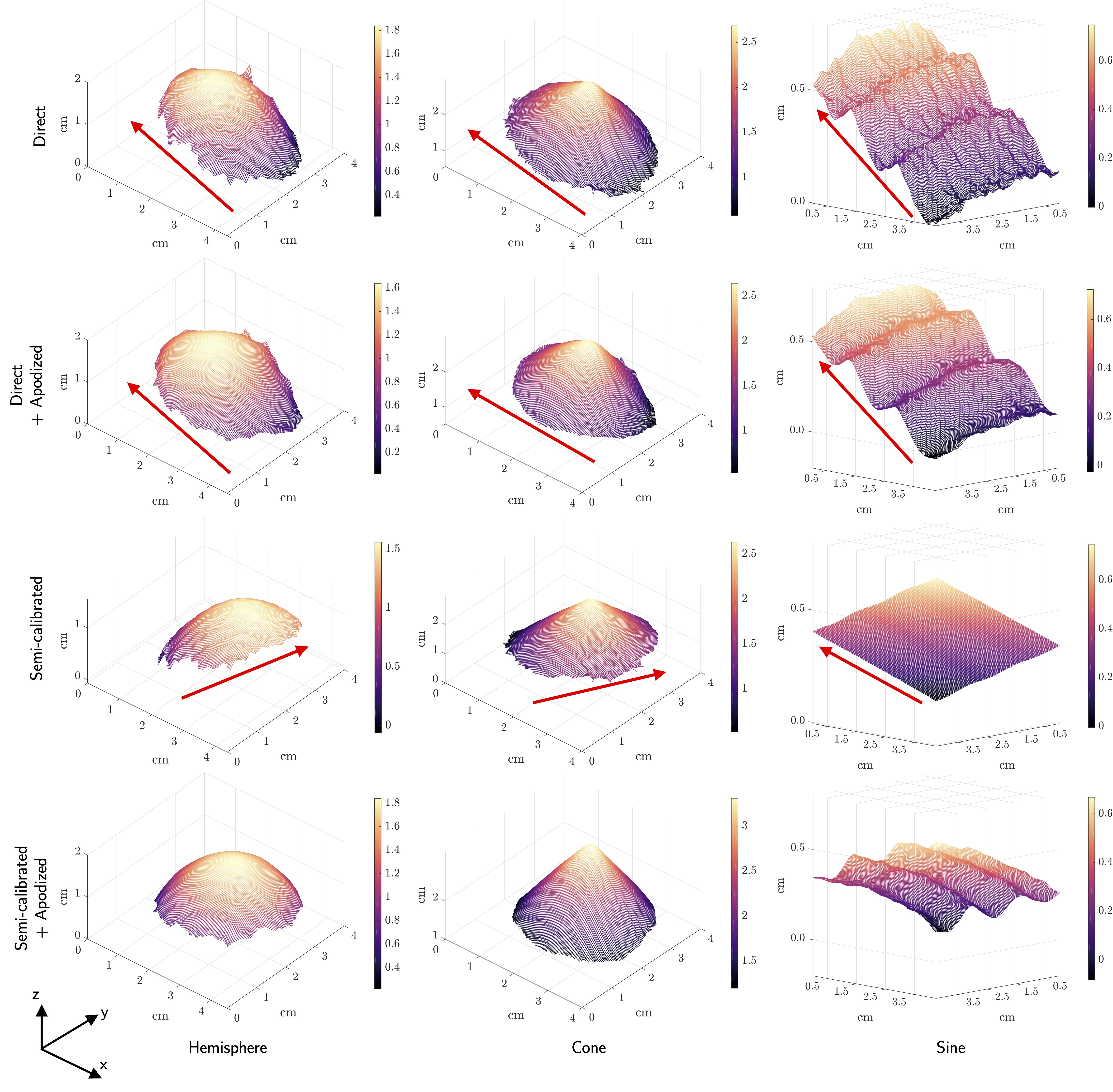}
		\caption{3D reconstructions of the hemisphere, cone, and sine wave objects under four different cases. Apart from the angular tilt for the direct photometric stereo cases (first two columns), the general 3D shape of the objects were more or less retrieved. The unapodized semi-calibrated case (third column), especially the sine wave object, has too much noise artifacts in the set of raw images. As a result, the non-linear least squares algorithm had difficulty looking for optimal solutions, yielding inaccurate 3D reconstructions. The apodized semi-calibrated case (fourth column) had the best results. Arrows were added to emphasize the tilt of the images.}
		\label{fig:3D}	
	\end{figure*}
	
	The spectra and reconstructed 2D images of the hemisphere object before and after Gaussian filtering from all six detectors are shown in Fig.~\ref{fig:spherescpectra}. 
	Additional sample images for the cone and sine wave object from one detector are shown in Fig.~\ref{fig:apodized}, with $\sigma$ for the Gaussian filter also set to be equal to $\alpha$.
	Each left-and-right pair corresponds to before-and-after apodizing the spectrum.
	Initially, the circular spectra have hard edges but after apodizing, the edges were softened out while features found in the lower spatial frequency bands were retained.
	An example of which is shown in Fig.~\ref{fig:sine3Dspectra} where the two prominent secondary peaks, whose distance from the central peak (DC term) represents the frequency of the sine wave object, were preserved.
	For a 2D sine wave, the amplitude of the DC term equals the average image intensity whereas the direction of the periodicity of the image feature is encoded in the orientation of the secondary peaks or the non-zero frequency components.
	In the case of our sine wave object, the secondary peaks we were able to obtain using our FGI setup are horizontally aligned (see Fig.~\ref{fig:apodized}h) with the central peak as a result of its vertical orientation (see Fig.~\ref{fig:objects}c) which agrees with Fourier theory.
	
	\begin{table*}[tbp]
		\centering
		\caption{Quantitative Analysis of the 3D Data$^\dagger$ and Intensity Error Maps$^\ddagger$ using Error Statistics.\footnote{The smallest value for each parameter under each object is highlighted. Smaller value means a more accurate estimated albedo-scaled normal. Due to inaccurate 3D result, the relative error for the semi-calibrated case of the sine wave was not included.}}
		\label{tab:errorestimates}
		\begin{ruledtabular}
			\begin{tabular}{l l c c c c c}
				\textbf{Object} & \textbf{Case} & \textbf{Relative Error}$^\dagger$ & \textbf{Angular Error}$^\dagger$ & \textbf{Mean}$^\ddagger$ & \textbf{Median}$^\ddagger$ & \textbf{Max}$^\ddagger$ \\
				\hline
				Hemisphere & Direct & 0.23 & $16.7^\circ$ & 0.049 & 0.053 & 0.12 \\
				& Direct + Apodized & 0.23 &  $10.8^\circ$ &  0.054 & 0.063 &	0.10 \\
				& Semi-calibrated & 0.40 & $20.5^\circ$ & 0.023 & 0.023 & 0.072 \\
				& Semi-calibrated + Apodized & \textbf{0.068} & \textbf{9.23}$^\circ$ & \textbf{0.012} & \textbf{0.012} & \textbf{0.034} \\ \hline
				Cone & Direct & 0.19 & $12.9^\circ$ & 0.052 & 0.053 & 0.13 \\
				& Direct + Apodized	& 0.18 & $8.73^\circ$ & 0.057 & 0.063 & 0.12 \\
				& Semi-calibrated & 0.24 & $13.9^\circ$ & 0.025	& 0.024 & 0.084 \\
				& Semi-calibrated + Apodized & \textbf{0.069} & \textbf{2.89}$^\circ$ & \textbf{0.013} & \textbf{0.012} & \textbf{0.037} \\ \hline
				Sine & Direct & 0.17 & $7.94^\circ$ & 0.072 & 0.069 & 0.18 \\
				& Direct + Apodized & 0.16 & $8.06^\circ$ & 0.071 & 0.071 & 0.16 \\
				& Semi-calibrated & $\times$ & $4.57^\circ$ & 0.054 & 0.051 & 0.15 \\
				& Semi-calibrated + Apodized & \textbf{0.065} & \textbf{4.01}$^\circ$ & \textbf{0.030} & \textbf{0.029} &\textbf{0.097} \\
			\end{tabular}
		\end{ruledtabular}
	\end{table*}
	
	Similar to the obtained results in our previous work \cite{Aguilar2019}, the signal-to-noise ratio is significantly reduced without a lens placed before the SPDs to collect the reflected light from the scene, resulting to grainy reconstructed images as seen in Figs.~\ref{fig:spherescpectra} and \ref{fig:apodized} especially before the apodization process.
	However, upon apodizing, the background noise present in the images was removed while the necessary lighting information, i.e. the varying illumination including the albedo of the object as well as shading effect based on the position of each detector, was otherwise retained.
	
	The SCPS algorithm is then used to obtain 3D reconstruction of the objects from the 2D images retrieved by the setup.
	For completeness, a direct 3D reconstruction without optimizing the lighting parameter by skipping Eq.~(\ref{eq:cost2}) in the reconstruction process in Section~\ref{sec:scps} is also performed.
	For this non-calibrated case, we assume the setup has fully-calibrated lighting implying the algorithm assumes the lighting strength $\gamma_i$ in Eq.~(\ref{eq:lightstrength}) has a unity value for each image.
	We will call this the direct photometric stereo (DPS) and we compare its result with that of SCPS for both unapodized and apodized cases.
	
	We provide the 3D reconstructions under each case for all our test objects (see Fig.~\ref{fig:3D}).
	For fair qualitative comparison, each set of 3D images for a particular object was given the same 3D view, i.e. the same azimuth and elevation settings (see \url{Visualization 2} for an animated 3D view of the cone for each case).
	Comparing first the unapodized with the apodized cases, the latter ones have little to no corrugation artifacts as a result of the softened Fourier spectrum edge while they are far more prominent in the former especially in the direct photometric stereo case for the sine wave.
	The results of the direct photometric stereo cases show inaccurate 3D reconstructions due to non-uniform lighting yielding tilted 3D images.
	
	\begin{figure}[b]
		\centering
		\includegraphics[width=0.95\linewidth]{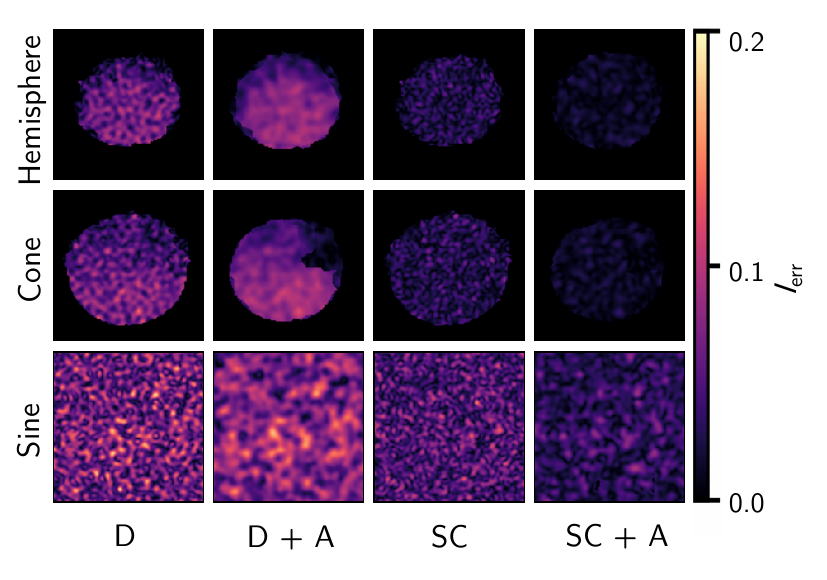}
		\caption{Intensity error maps for each case--object combination.  The symbols `D' and `SC' denote direct and semi-calibrated photometric stereo cases, respectively, while `+A' denotes apodized 2D results. Each pixel corresponds to the sum of root-mean-square errors over the set of images captured by the six photoresistors at the same pixel location.}
		\label{fig:intensityerrormaps}	
	\end{figure}
	
	The 3D results were evaluated quantitatively by obtaining the relative error with respect to the ground truth.
	The parameter we considered for the depth relative error calculation was the radius for the hemisphere, height for the cone, and wavelength for the sine wave object with the estimated measurements given in Sec.~\ref{sec:experiment}.
	3D reconstruction errors are typically expressed as the root-mean-square error (RMSE) between a precise 3D measurement of the surface, e.g. by laser ranging, and the proposed technique. 
	Without a laser ranger, we instead compared of the objects at reference points to their corresponding 3D measurements retrieved using the algorithm.
	The angular errors of the objects were also calculated to quantify the tilting based on nonuniform exposure.
	The results are tabulated in Tab.~\ref{tab:errorestimates}.
	The high values for the relative errors are mainly due to the corrugation artifacts and tilt distortions.
	For the unapodized semi-calibrated cases, especially for the sine wave, the intensity matrices were badly scaled as a consequence of too much noise artifacts in the set of 2D unapodized images. 
	This led to inaccurate results as the non-linear least squares algorithm had difficulty in finding best optimal solutions which proves to be a limitation of SCPS.
	For the apodized semi-calibrated cases, errors were reduced but are still present.
	These errors occur near the edges, tips, and great depth transitions. 
	If apodizing effect is ignored, as was pointed out in Ref.~\cite{Sun2013}, a limitation of 3D CGI systems would be the distortions caused by the expanding projected patterns at greater depths but may be slightly corrected with a projection lens of suitable focal length.
	
	For all objects, the apodized semi-calibrated cases have the lowest relative error, achieving at least 93.5\% accuracy. 
	They also have the lowest angular error.
	This proves the effectiveness of using the this method for the 3D FGI system.
	We take the apodized hemisphere images as an example.
	Optimized lighting parameters by using the cost function minimization algorithm in Eqs.~(\ref{eq:cost1})-(\ref{eq:cost2}) yielded the adjusted lighting strength matrix $\mathbf{\Gamma} = \left[1.1, 1.3, 0.72, 0.94, 1.1, 0.78\right]$.
	Even if the photoresistors have different sensitivities leading to light strength variations in the 2D images, errors upon 3D reconstruction due to non-uniform intensities received by the detectors are corrected using SCPS while important details are preserved.
	
	We also evaluate the estimated albedo-scaled surface normal $\mathbf{g}$, which dictates the resulting 3D shape for each case, by plugging it back to Eq.~(\ref{eq:psintensitymatrix}) and calculating the intensity error 
	\begin{equation}
	I_\text{err} = \sum_{i=1}^M \sqrt{\dfrac{\sum_{k=0}^{K-1}\sum_{l=0}^{L-1}(I_i(k,l) - I_{\text{est},i}(k,l))^2}{K*L}},    
	\end{equation}
	where $I_{\text{est},i} = \mathbf{d}_i^\top \mathbf{g}$ with $\textbf{d}_i$ being either an unoptimized or optimized lighting parameter for the DPS or SCPS case, respectively.
	The overall intensity variation across the different images is obtained.
	In particular, for one object, we calculate the RMSE at each pixel location across the set of images retrieved by the six photoresistors. 
	The intensity error map has the same size as the images and pixel values ranging from [0,1].
	Figure~\ref{fig:intensityerrormaps} shows the resulting intensity error maps with the apodized semi-calibrated (SC + A) case having the least intensity variation followed by the unapodized semi-calibrated (SC), unapodized direct (D), and apodized direct (D + A) case for each object.
	We apply error statistics to further evaluate each error map by calculating the mean, median, and maximum value of all the points in the map.
	By looking at these three parameters, we can have a fair quantitative comparison of the goodness of reconstruction of the four cases. 
	
	The calculated error statistics are also shown in Tab.~\ref{tab:errorestimates}.
	A value closer to zero for each parameter means a more accurate estimated albedo-scaled surface normal $\mathbf{g}$. 
	For each object, the cases in which we directly applied photometric stereo without optimizing light parameters (D and D + A) almost have the same error statistics.
	The SC + A cases have the smallest mean error as well as median and maximum values followed by the unapodized SC cases.
	Among the three objects, the sine wave has the highest error statistics with a mean intensity error of 0.03.
	However, this still gives us at least 97.0\% similarity between the estimated and the actual value for the SC + A case followed by the SC, D + A, and D case at 94.6\%, 92.9\%, and 92.8\%, respectively.
	Although the error maps of the unapodized semi-calibrated cases have better results compared to both apodized and unapodized direct cases, it does not necessarily translate to more accurate 3D result considering the noise artifacts have greatly affected the optimization algorithm.
	
	Both qualitative and quantitative analysis of the 3D results and the estimated surface normals using error statistics have shown apodized SCPS cases outperform the others as evident with the more accurate results.
	From the statistics of the intensity error maps of the three objects, we see that the hemisphere has the most accurate estimated albedo-scaled surface normal followed by the cone and then the sine wave.
	Apart from the simpler structure of the hemisphere and cone relative to the sine wave, this may be attributed to the use of shadow masks for these two objects. 
	We masked out their backgrounds which effectively helped the algorithm identify and isolate the actual object in the images.
	For the sine wave, the images were directly fed to the photometric stereo algorithm without masking since the entire illuminated scene constituted the object.
	
	Our 3D FGI system was able to yield satisfying results even at 5\% spectral coverage ratio. 
	Furthermore, with sub-Nyquist sampling in the Fourier domain, far fewer measurements are needed for our algorithm as compared with Ref.~\cite{Sun2013} or the several 3D CGI techniques presented in Ref.~\cite{Sun2019}.
	The artifacts in the 2D images carried out in the 3D reconstruction were primarily caused by the absence of a focusing lens.
	However, other factors may include voltage fluctuations from analog-to-digital conversion, ambient light entering the experimental setup, photoresistor thermal noise, and pattern degradation as it reaches the scene due to environmental noise or projection quality of the commercial projector.
	These problems and limitations may be addressed by averaging over multiple trials if time is not a constraint, or in general, improving upon the environment of the experimental setup and other elements needed for imaging.
	
	\section{Conclusion and Recommendations}
	We performed 3D Fourier ghost imaging using semi-calibrated photometric stereo with photoresistors as single-pixel detectors and a commercial projector for light pattern projection. 
	The results of the SCPS algorithm were compared with that of DPS for both apodized and unapodized cases.
	Qualitative and quantitative analyses of the 3D reconstructed results of three test objects (sphere, cone, 3D sine wave) showed the apodized SCPS case has the most accurate 3D reconstructions, achieving a 93.5\% accuracy.
	Moreover, intensity error map statistics yielded a 97\% accuracy for the estimated surface normals using SCPS.
	
	While the SCPS algorithm can correct or minimize the angular tilt in the 3D images due to non-uniform lighting, too much noise artifacts can cause the optimization technique to produce inaccurate 3D result as evident in the unapodized SCPS cases.
	To avoid this, other noise filtering techniques can be applied to the 2D images before feeding them to the algorithm but not to the point of removing necessary details in the object or altering its overall structure.
	The extent of apodization can also be further explored for more complicated objects, i.e., choosing an appropriate value for the radius of the Gaussian filter based on the object's spatial resolution.
	If post-processing is not an option, averaging over multiple trials can also reduce the noise artifacts but will lead to longer experiment time.
	
	The few samples needed (at 5\% sub-Nyquist sampling frequency) to reconstruct an image makes the imaging system capable of real-time 3D imaging of scenes without complex structure or details.
	Furthermore, the system can be still improved if SPDs and projectors with faster acquisition and projection time, respectively, were used. 
	In itself, the system is already cheap and adaptable even with the use of six photoresistors that can be further reduced to a minimum of three. 
	The system has the potential to be used and extended to wavelengths beyond the visible light at a cheaper cost than conventional imaging.
	
	\begin{description}
		\item[Funding] National Research Council of the Philippines (NRCP).
		\item[Disclosures] The authors declare no conflicts of interest.
		\item[Data availability] Data underlying the results presented in this paper are not publicly available at this time but may be obtained from the authors upon reasonable request.
	\end{description}
	
	
	
	
	
	\bibliography{apssamp}

\begin{thebibliography}{56}%
\makeatletter
\providecommand \@ifxundefined [1]{%
 \@ifx{#1\undefined}
}%
\providecommand \@ifnum [1]{%
 \ifnum #1\expandafter \@firstoftwo
 \else \expandafter \@secondoftwo
 \fi
}%
\providecommand \@ifx [1]{%
 \ifx #1\expandafter \@firstoftwo
 \else \expandafter \@secondoftwo
 \fi
}%
\providecommand \natexlab [1]{#1}%
\providecommand \enquote  [1]{``#1''}%
\providecommand \bibnamefont  [1]{#1}%
\providecommand \bibfnamefont [1]{#1}%
\providecommand \citenamefont [1]{#1}%
\providecommand \href@noop [0]{\@secondoftwo}%
\providecommand \href [0]{\begingroup \@sanitize@url \@href}%
\providecommand \@href[1]{\@@startlink{#1}\@@href}%
\providecommand \@@href[1]{\endgroup#1\@@endlink}%
\providecommand \@sanitize@url [0]{\catcode `\\12\catcode `\$12\catcode
  `\&12\catcode `\#12\catcode `\^12\catcode `\_12\catcode `\%12\relax}%
\providecommand \@@startlink[1]{}%
\providecommand \@@endlink[0]{}%
\providecommand \url  [0]{\begingroup\@sanitize@url \@url }%
\providecommand \@url [1]{\endgroup\@href {#1}{\urlprefix }}%
\providecommand \urlprefix  [0]{URL }%
\providecommand \Eprint [0]{\href }%
\providecommand \doibase [0]{https://doi.org/}%
\providecommand \selectlanguage [0]{\@gobble}%
\providecommand \bibinfo  [0]{\@secondoftwo}%
\providecommand \bibfield  [0]{\@secondoftwo}%
\providecommand \translation [1]{[#1]}%
\providecommand \BibitemOpen [0]{}%
\providecommand \bibitemStop [0]{}%
\providecommand \bibitemNoStop [0]{.\EOS\space}%
\providecommand \EOS [0]{\spacefactor3000\relax}%
\providecommand \BibitemShut  [1]{\csname bibitem#1\endcsname}%
\let\auto@bib@innerbib\@empty
\bibitem [{\citenamefont {Pittman}\ \emph {et~al.}(1995)\citenamefont
  {Pittman}, \citenamefont {Shih}, \citenamefont {Strekalov},\ and\
  \citenamefont {Sergienko}}]{Pittman1995}%
  \BibitemOpen
  \bibfield  {author} {\bibinfo {author} {\bibfnamefont {T.~B.}\ \bibnamefont
  {Pittman}}, \bibinfo {author} {\bibfnamefont {Y.~H.}\ \bibnamefont {Shih}},
  \bibinfo {author} {\bibfnamefont {D.~V.}\ \bibnamefont {Strekalov}},\ and\
  \bibinfo {author} {\bibfnamefont {A.~V.}\ \bibnamefont {Sergienko}},\
  }\bibfield  {title} {\bibinfo {title} {{Optical imaging by means of
  two-photon quantum entanglement}},\ }\href
  {https://doi.org/10.1103/PhysRevA.52.R3429} {\bibfield  {journal} {\bibinfo
  {journal} {Phys. Rev. A}\ }\textbf {\bibinfo {volume} {52}},\ \bibinfo
  {pages} {R3429} (\bibinfo {year} {1995})}\BibitemShut {NoStop}%
\bibitem [{\citenamefont {Bennink}\ \emph {et~al.}(2002)\citenamefont
  {Bennink}, \citenamefont {Bentley},\ and\ \citenamefont
  {Boyd}}]{Bennink2002}%
  \BibitemOpen
  \bibfield  {author} {\bibinfo {author} {\bibfnamefont {R.~S.}\ \bibnamefont
  {Bennink}}, \bibinfo {author} {\bibfnamefont {S.~J.}\ \bibnamefont
  {Bentley}},\ and\ \bibinfo {author} {\bibfnamefont {R.~W.}\ \bibnamefont
  {Boyd}},\ }\bibfield  {title} {\bibinfo {title} {{Two-Photon” Coincidence
  Imaging with a Classical Source}},\ }\href
  {https://doi.org/10.1103/PhysRevLett.89.113601} {\bibfield  {journal}
  {\bibinfo  {journal} {Phys. Rev. Lett.}\ }\textbf {\bibinfo {volume} {89}},\
  \bibinfo {pages} {113601} (\bibinfo {year} {2002})}\BibitemShut {NoStop}%
\bibitem [{\citenamefont {Valencia}\ \emph {et~al.}(2005)\citenamefont
  {Valencia}, \citenamefont {Scarcelli}, \citenamefont {D'Angelo},\ and\
  \citenamefont {Shih}}]{Valencia2005}%
  \BibitemOpen
  \bibfield  {author} {\bibinfo {author} {\bibfnamefont {A.}~\bibnamefont
  {Valencia}}, \bibinfo {author} {\bibfnamefont {G.}~\bibnamefont {Scarcelli}},
  \bibinfo {author} {\bibfnamefont {M.}~\bibnamefont {D'Angelo}},\ and\
  \bibinfo {author} {\bibfnamefont {Y.}~\bibnamefont {Shih}},\ }\bibfield
  {title} {\bibinfo {title} {{Two-Photon Imaging with Thermal Light}},\ }\href
  {https://doi.org/10.1103/PhysRevLett.94.063601} {\bibfield  {journal}
  {\bibinfo  {journal} {Phys. Rev. Lett.}\ }\textbf {\bibinfo {volume} {94}},\
  \bibinfo {pages} {063601} (\bibinfo {year} {2005})}\BibitemShut {NoStop}%
\bibitem [{\citenamefont {Ferri}\ \emph {et~al.}(2005)\citenamefont {Ferri},
  \citenamefont {Magatti}, \citenamefont {Gatti}, \citenamefont {Bache},
  \citenamefont {Brambilla},\ and\ \citenamefont {Lugiato}}]{Ferri2005}%
  \BibitemOpen
  \bibfield  {author} {\bibinfo {author} {\bibfnamefont {F.}~\bibnamefont
  {Ferri}}, \bibinfo {author} {\bibfnamefont {D.}~\bibnamefont {Magatti}},
  \bibinfo {author} {\bibfnamefont {A.}~\bibnamefont {Gatti}}, \bibinfo
  {author} {\bibfnamefont {M.}~\bibnamefont {Bache}}, \bibinfo {author}
  {\bibfnamefont {E.}~\bibnamefont {Brambilla}},\ and\ \bibinfo {author}
  {\bibfnamefont {L.~A.}\ \bibnamefont {Lugiato}},\ }\bibfield  {title}
  {\bibinfo {title} {{High-resolution ghost image and ghost diffraction
  experiments with thermal light}},\ }\href
  {https://doi.org/10.1103/PhysRevLett.94.183602} {\bibfield  {journal}
  {\bibinfo  {journal} {Phys. Rev. Lett.}\ }\textbf {\bibinfo {volume} {94}},\
  \bibinfo {pages} {183602} (\bibinfo {year} {2005})}\BibitemShut {NoStop}%
\bibitem [{\citenamefont {Zhang}\ \emph {et~al.}(2014)\citenamefont {Zhang},
  \citenamefont {Tang}, \citenamefont {Wu}, \citenamefont {Qiu}, \citenamefont
  {Xu}, \citenamefont {Li}, \citenamefont {Wang}, \citenamefont {Xiong},\ and\
  \citenamefont {Wang}}]{Zhang2014}%
  \BibitemOpen
  \bibfield  {author} {\bibinfo {author} {\bibfnamefont {D.-J.}\ \bibnamefont
  {Zhang}}, \bibinfo {author} {\bibfnamefont {Q.}~\bibnamefont {Tang}},
  \bibinfo {author} {\bibfnamefont {T.-F.}\ \bibnamefont {Wu}}, \bibinfo
  {author} {\bibfnamefont {H.-C.}\ \bibnamefont {Qiu}}, \bibinfo {author}
  {\bibfnamefont {D.-Q.}\ \bibnamefont {Xu}}, \bibinfo {author} {\bibfnamefont
  {H.-G.}\ \bibnamefont {Li}}, \bibinfo {author} {\bibfnamefont {H.-B.}\
  \bibnamefont {Wang}}, \bibinfo {author} {\bibfnamefont {J.}~\bibnamefont
  {Xiong}},\ and\ \bibinfo {author} {\bibfnamefont {K.}~\bibnamefont {Wang}},\
  }\bibfield  {title} {\bibinfo {title} {{Lensless ghost imaging of a phase
  object with pseudo-thermal light}},\ }\href
  {https://doi.org/10.1063/1.4869959} {\bibfield  {journal} {\bibinfo
  {journal} {App. Phys. Lett.}\ }\textbf {\bibinfo {volume} {104}},\ \bibinfo
  {pages} {121113} (\bibinfo {year} {2014})}\BibitemShut {NoStop}%
\bibitem [{\citenamefont {Devaux}\ \emph {et~al.}(2017)\citenamefont {Devaux},
  \citenamefont {Huy}, \citenamefont {Denis}, \citenamefont {Lantz},\ and\
  \citenamefont {Moreau}}]{Devaux2017}%
  \BibitemOpen
  \bibfield  {author} {\bibinfo {author} {\bibfnamefont {F.}~\bibnamefont
  {Devaux}}, \bibinfo {author} {\bibfnamefont {K.~P.}\ \bibnamefont {Huy}},
  \bibinfo {author} {\bibfnamefont {S.}~\bibnamefont {Denis}}, \bibinfo
  {author} {\bibfnamefont {E.}~\bibnamefont {Lantz}},\ and\ \bibinfo {author}
  {\bibfnamefont {P.-A.}\ \bibnamefont {Moreau}},\ }\bibfield  {title}
  {\bibinfo {title} {{Temporal ghost imaging with pseudo-thermal speckle
  light}},\ }\href {https://doi.org/10.1088/2040-8986/aa5328} {\bibfield
  {journal} {\bibinfo  {journal} {J. Opt.}\ }\textbf {\bibinfo {volume} {19}},\
  \bibinfo {pages} {024001} (\bibinfo {year} {2017})}\BibitemShut {NoStop}%
\bibitem [{\citenamefont {Bennink}\ \emph {et~al.}(2004)\citenamefont
  {Bennink}, \citenamefont {Bentley}, \citenamefont {Boyd},\ and\ \citenamefont
  {Howell}}]{Bennink2004}%
  \BibitemOpen
  \bibfield  {author} {\bibinfo {author} {\bibfnamefont {R.~S.}\ \bibnamefont
  {Bennink}}, \bibinfo {author} {\bibfnamefont {S.~J.}\ \bibnamefont
  {Bentley}}, \bibinfo {author} {\bibfnamefont {R.~W.}\ \bibnamefont {Boyd}},\
  and\ \bibinfo {author} {\bibfnamefont {J.~C.}\ \bibnamefont {Howell}},\
  }\bibfield  {title} {\bibinfo {title} {{Quantum and Classical Coincidence
  Imaging}},\ }\href {https://doi.org/10.1103/PhysRevLett.92.033601} {\bibfield
   {journal} {\bibinfo  {journal} {Phys. Rev. Lett.}\ }\textbf {\bibinfo
  {volume} {92}},\ \bibinfo {pages} {033601} (\bibinfo {year}
  {2004})}\BibitemShut {NoStop}%
\bibitem [{\citenamefont {Gatti}\ \emph {et~al.}(2004)\citenamefont {Gatti},
  \citenamefont {Brambilla}, \citenamefont {Bache},\ and\ \citenamefont
  {Lugiato}}]{Gatti2004}%
  \BibitemOpen
  \bibfield  {author} {\bibinfo {author} {\bibfnamefont {A.}~\bibnamefont
  {Gatti}}, \bibinfo {author} {\bibfnamefont {E.}~\bibnamefont {Brambilla}},
  \bibinfo {author} {\bibfnamefont {M.}~\bibnamefont {Bache}},\ and\ \bibinfo
  {author} {\bibfnamefont {L.~A.}\ \bibnamefont {Lugiato}},\ }\bibfield
  {title} {\bibinfo {title} {{Ghost Imaging with Thermal Light: Comparing
  Entanglement and Classical Correlation}},\ }\href
  {https://doi.org/10.1103/PhysRevLett.93.093602} {\bibfield  {journal}
  {\bibinfo  {journal} {Phys. Rev. Lett.}\ }\textbf {\bibinfo {volume} {93}},\
  \bibinfo {pages} {093602} (\bibinfo {year} {2004})}\BibitemShut {NoStop}%
\bibitem [{\citenamefont {Shapiro}(2008)}]{Shapiro2008}%
  \BibitemOpen
  \bibfield  {author} {\bibinfo {author} {\bibfnamefont {J.~H.}\ \bibnamefont
  {Shapiro}},\ }\bibfield  {title} {\bibinfo {title} {Computational ghost
  imaging},\ }\href {https://doi.org/10.1103/PhysRevA.78.061802} {\bibfield
  {journal} {\bibinfo  {journal} {Phys. Rev. A}\ }\textbf {\bibinfo {volume}
  {78}},\ \bibinfo {pages} {061802} (\bibinfo {year} {2008})}\BibitemShut
  {NoStop}%
\bibitem [{\citenamefont {Cand{\`{e}}s}(2006)}]{Candes2006}%
  \BibitemOpen
  \bibfield  {author} {\bibinfo {author} {\bibfnamefont {E.~J.}\ \bibnamefont
  {Cand{\`{e}}s}},\ }\bibfield  {title} {\bibinfo {title} {{Compressive
  sampling}},\ }in\ \href
  {http://citeseerx.ist.psu.edu/viewdoc/download?doi=10.1.1.419.4969{\&}rep=rep1{\&}type=pdf}
  {\emph {\bibinfo {booktitle} {Proc. Int. Con. Math.}}}\ (\bibinfo {year}
  {2006})\BibitemShut {NoStop}%
\bibitem [{\citenamefont {Donoho}(2006)}]{Donoho2006}%
  \BibitemOpen
  \bibfield  {author} {\bibinfo {author} {\bibfnamefont {D.~L.}\ \bibnamefont
  {Donoho}},\ }\bibfield  {title} {\bibinfo {title} {{Compressed Sensing}},\
  }\href {https://doi.org/10.1109/TIT.2006.871582} {\bibfield  {journal}
  {\bibinfo  {journal} {IEEE Trans. Inf. Theory}\ }\textbf {\bibinfo {volume}
  {52}},\ \bibinfo {pages} {1289} (\bibinfo {year} {2006})}\BibitemShut
  {NoStop}%
\bibitem [{\citenamefont {Duarte}\ \emph {et~al.}(2008)\citenamefont {Duarte},
  \citenamefont {Davenport}, \citenamefont {Takbar}, \citenamefont {Laska},
  \citenamefont {Sun}, \citenamefont {Kelly},\ and\ \citenamefont
  {Baraniuk}}]{Duarte2008a}%
  \BibitemOpen
  \bibfield  {author} {\bibinfo {author} {\bibfnamefont {M.~F.}\ \bibnamefont
  {Duarte}}, \bibinfo {author} {\bibfnamefont {M.~A.}\ \bibnamefont
  {Davenport}}, \bibinfo {author} {\bibfnamefont {D.}~\bibnamefont {Takbar}},
  \bibinfo {author} {\bibfnamefont {J.~N.}\ \bibnamefont {Laska}}, \bibinfo
  {author} {\bibfnamefont {T.}~\bibnamefont {Sun}}, \bibinfo {author}
  {\bibfnamefont {K.~F.}\ \bibnamefont {Kelly}},\ and\ \bibinfo {author}
  {\bibfnamefont {R.~G.}\ \bibnamefont {Baraniuk}},\ }\bibfield  {title}
  {\bibinfo {title} {{Single-pixel imaging via compressive sampling: Building
  simpler, smaller, and less-expensive digital cameras}},\ }\href
  {https://doi.org/10.1109/MSP.2007.914730} {\bibfield  {journal} {\bibinfo
  {journal} {IEEE Signal Process. Mag.}\ }\textbf {\bibinfo {volume} {25}},\
  \bibinfo {pages} {83} (\bibinfo {year} {2008})}\BibitemShut {NoStop}%
\bibitem [{\citenamefont {Edgar}\ \emph {et~al.}(2015)\citenamefont {Edgar},
  \citenamefont {Gibson}, \citenamefont {Bowman}, \citenamefont {Sun},
  \citenamefont {Radwell}, \citenamefont {Mitchell}, \citenamefont {Welsh},\
  and\ \citenamefont {Padgett}}]{Edgar2015_1}%
  \BibitemOpen
  \bibfield  {author} {\bibinfo {author} {\bibfnamefont {M.~P.}\ \bibnamefont
  {Edgar}}, \bibinfo {author} {\bibfnamefont {G.~M.}\ \bibnamefont {Gibson}},
  \bibinfo {author} {\bibfnamefont {R.~W.}\ \bibnamefont {Bowman}}, \bibinfo
  {author} {\bibfnamefont {B.}~\bibnamefont {Sun}}, \bibinfo {author}
  {\bibfnamefont {N.}~\bibnamefont {Radwell}}, \bibinfo {author} {\bibfnamefont
  {K.~J.}\ \bibnamefont {Mitchell}}, \bibinfo {author} {\bibfnamefont {S.~S.}\
  \bibnamefont {Welsh}},\ and\ \bibinfo {author} {\bibfnamefont {M.~J.}\
  \bibnamefont {Padgett}},\ }\bibfield  {title} {\bibinfo {title}
  {{Simultaneous real-time visible and infrared video with single-pixel
  detectors}},\ }\href {https://doi.org/10.1038/srep10669} {\bibfield
  {journal} {\bibinfo  {journal} {Sci. Rep.}\ }\textbf {\bibinfo {volume}
  {5}},\ \bibinfo {pages} {10669} (\bibinfo {year} {2015})}\BibitemShut
  {NoStop}%
\bibitem [{\citenamefont {Stantchev}\ \emph {et~al.}(2016)\citenamefont
  {Stantchev}, \citenamefont {Sun}, \citenamefont {Hornett}, \citenamefont
  {Hobson}, \citenamefont {Gibson}, \citenamefont {Padgett},\ and\
  \citenamefont {Hendry}}]{Stantchev2016}%
  \BibitemOpen
  \bibfield  {author} {\bibinfo {author} {\bibfnamefont {R.~I.}\ \bibnamefont
  {Stantchev}}, \bibinfo {author} {\bibfnamefont {B.}~\bibnamefont {Sun}},
  \bibinfo {author} {\bibfnamefont {S.~M.}\ \bibnamefont {Hornett}}, \bibinfo
  {author} {\bibfnamefont {P.~A.}\ \bibnamefont {Hobson}}, \bibinfo {author}
  {\bibfnamefont {G.~M.}\ \bibnamefont {Gibson}}, \bibinfo {author}
  {\bibfnamefont {M.~J.}\ \bibnamefont {Padgett}},\ and\ \bibinfo {author}
  {\bibfnamefont {E.}~\bibnamefont {Hendry}},\ }\bibfield  {title} {\bibinfo
  {title} {{Noninvasive, near-field terahertz imaging of hidden objects using a
  single-pixel detector}},\ }\bibfield  {journal} {\bibinfo  {journal} {Science
  Advances}\ }\textbf {\bibinfo {volume} {2}},\ \href
  {https://doi.org/10.1126/sciadv.1600190} {10.1126/sciadv.1600190} (\bibinfo
  {year} {2016})\BibitemShut {NoStop}%
\bibitem [{\citenamefont {Gibson}\ \emph {et~al.}(2017)\citenamefont {Gibson},
  \citenamefont {Sun}, \citenamefont {Edgar}, \citenamefont {Phillips},
  \citenamefont {Hempler}, \citenamefont {Maker}, \citenamefont {Malcolm},\
  and\ \citenamefont {Padgett}}]{Gibson2017}%
  \BibitemOpen
  \bibfield  {author} {\bibinfo {author} {\bibfnamefont {G.~M.}\ \bibnamefont
  {Gibson}}, \bibinfo {author} {\bibfnamefont {B.}~\bibnamefont {Sun}},
  \bibinfo {author} {\bibfnamefont {M.~P.}\ \bibnamefont {Edgar}}, \bibinfo
  {author} {\bibfnamefont {D.~B.}\ \bibnamefont {Phillips}}, \bibinfo {author}
  {\bibfnamefont {N.}~\bibnamefont {Hempler}}, \bibinfo {author} {\bibfnamefont
  {G.~T.}\ \bibnamefont {Maker}}, \bibinfo {author} {\bibfnamefont {G.~P.~A.}\
  \bibnamefont {Malcolm}},\ and\ \bibinfo {author} {\bibfnamefont {M.~J.}\
  \bibnamefont {Padgett}},\ }\bibfield  {title} {\bibinfo {title} {{Real-time
  imaging of methane gas leaks using a single-pixel camera}},\ }\href
  {https://doi.org/10.1364/OE.25.002998} {\bibfield  {journal} {\bibinfo
  {journal} {Opt. Exp.}\ }\textbf {\bibinfo {volume} {25}},\ \bibinfo {pages}
  {2998} (\bibinfo {year} {2017})}\BibitemShut {NoStop}%
\bibitem [{\citenamefont {Kim}\ \emph {et~al.}(2020)\citenamefont {Kim},
  \citenamefont {Gelisio}, \citenamefont {Mercurio}, \citenamefont
  {Dziarzhytski}, \citenamefont {Beye}, \citenamefont {Bocklage}, \citenamefont
  {Classen}, \citenamefont {David}, \citenamefont {Gorobtsov}, \citenamefont
  {Khubbutdinov}, \citenamefont {Lazarev}, \citenamefont {Mukharamova},
  \citenamefont {Obukhov}, \citenamefont {R\"osner}, \citenamefont {Schlage},
  \citenamefont {Zaluzhnyy}, \citenamefont {Brenner}, \citenamefont
  {R\"ohlsberger}, \citenamefont {von Zanthier}, \citenamefont {Wurth},\ and\
  \citenamefont {Vartanyants}}]{Kim2020}%
  \BibitemOpen
  \bibfield  {author} {\bibinfo {author} {\bibfnamefont {Y.~Y.}\ \bibnamefont
  {Kim}}, \bibinfo {author} {\bibfnamefont {L.}~\bibnamefont {Gelisio}},
  \bibinfo {author} {\bibfnamefont {G.}~\bibnamefont {Mercurio}}, \bibinfo
  {author} {\bibfnamefont {S.}~\bibnamefont {Dziarzhytski}}, \bibinfo {author}
  {\bibfnamefont {M.}~\bibnamefont {Beye}}, \bibinfo {author} {\bibfnamefont
  {L.}~\bibnamefont {Bocklage}}, \bibinfo {author} {\bibfnamefont
  {A.}~\bibnamefont {Classen}}, \bibinfo {author} {\bibfnamefont
  {C.}~\bibnamefont {David}}, \bibinfo {author} {\bibfnamefont {O.~Y.}\
  \bibnamefont {Gorobtsov}}, \bibinfo {author} {\bibfnamefont {R.}~\bibnamefont
  {Khubbutdinov}}, \bibinfo {author} {\bibfnamefont {S.}~\bibnamefont
  {Lazarev}}, \bibinfo {author} {\bibfnamefont {N.}~\bibnamefont
  {Mukharamova}}, \bibinfo {author} {\bibfnamefont {Y.~N.}\ \bibnamefont
  {Obukhov}}, \bibinfo {author} {\bibfnamefont {B.}~\bibnamefont {R\"osner}},
  \bibinfo {author} {\bibfnamefont {K.}~\bibnamefont {Schlage}}, \bibinfo
  {author} {\bibfnamefont {I.~A.}\ \bibnamefont {Zaluzhnyy}}, \bibinfo {author}
  {\bibfnamefont {G.}~\bibnamefont {Brenner}}, \bibinfo {author} {\bibfnamefont
  {R.}~\bibnamefont {R\"ohlsberger}}, \bibinfo {author} {\bibfnamefont
  {J.}~\bibnamefont {von Zanthier}}, \bibinfo {author} {\bibfnamefont
  {W.}~\bibnamefont {Wurth}},\ and\ \bibinfo {author} {\bibfnamefont {I.~A.}\
  \bibnamefont {Vartanyants}},\ }\bibfield  {title} {\bibinfo {title} {Ghost
  imaging at an \text{XUV} free-electron laser},\ }\href
  {https://doi.org/10.1103/PhysRevA.101.013820} {\bibfield  {journal} {\bibinfo
   {journal} {Phys. Rev. A}\ }\textbf {\bibinfo {volume} {101}},\ \bibinfo
  {pages} {013820} (\bibinfo {year} {2020})}\BibitemShut {NoStop}%
\bibitem [{\citenamefont {Kingston}\ \emph {et~al.}(2018)\citenamefont
  {Kingston}, \citenamefont {Pelliccia}, \citenamefont {Rack}, \citenamefont
  {Olbinado}, \citenamefont {Cheng}, \citenamefont {Myers},\ and\ \citenamefont
  {Paganin}}]{Kingston2018}%
  \BibitemOpen
  \bibfield  {author} {\bibinfo {author} {\bibfnamefont {A.~M.}\ \bibnamefont
  {Kingston}}, \bibinfo {author} {\bibfnamefont {D.}~\bibnamefont {Pelliccia}},
  \bibinfo {author} {\bibfnamefont {A.}~\bibnamefont {Rack}}, \bibinfo {author}
  {\bibfnamefont {M.~P.}\ \bibnamefont {Olbinado}}, \bibinfo {author}
  {\bibfnamefont {Y.}~\bibnamefont {Cheng}}, \bibinfo {author} {\bibfnamefont
  {G.~R.}\ \bibnamefont {Myers}},\ and\ \bibinfo {author} {\bibfnamefont
  {D.~M.}\ \bibnamefont {Paganin}},\ }\bibfield  {title} {\bibinfo {title}
  {{Ghost tomography}},\ }\href {https://doi.org/10.1364/OPTICA.5.001516}
  {\bibfield  {journal} {\bibinfo  {journal} {Optica}\ }\textbf {\bibinfo
  {volume} {5}},\ \bibinfo {pages} {1516} (\bibinfo {year} {2018})}\BibitemShut
  {NoStop}%
\bibitem [{\citenamefont {Zhang}\ \emph
  {et~al.}(2018{\natexlab{a}})\citenamefont {Zhang}, \citenamefont {He},
  \citenamefont {Wu}, \citenamefont {Chen},\ and\ \citenamefont
  {Wang}}]{Zhang2018a}%
  \BibitemOpen
  \bibfield  {author} {\bibinfo {author} {\bibfnamefont {A.-X.}\ \bibnamefont
  {Zhang}}, \bibinfo {author} {\bibfnamefont {Y.-H.}\ \bibnamefont {He}},
  \bibinfo {author} {\bibfnamefont {L.-A.}\ \bibnamefont {Wu}}, \bibinfo
  {author} {\bibfnamefont {L.-M.}\ \bibnamefont {Chen}},\ and\ \bibinfo
  {author} {\bibfnamefont {B.-B.}\ \bibnamefont {Wang}},\ }\bibfield  {title}
  {\bibinfo {title} {{Tabletop x-ray ghost imaging with ultra-low radiation}},\
  }\href {https://doi.org/10.1364/OPTICA.5.000374} {\bibfield  {journal}
  {\bibinfo  {journal} {Optica}\ }\textbf {\bibinfo {volume} {5}},\ \bibinfo
  {pages} {374} (\bibinfo {year} {2018}{\natexlab{a}})}\BibitemShut {NoStop}%
\bibitem [{\citenamefont {Scarcelli}\ \emph {et~al.}(2003)\citenamefont
  {Scarcelli}, \citenamefont {Valencia}, \citenamefont {Gompers},\ and\
  \citenamefont {Shih}}]{Scarcelli2003}%
  \BibitemOpen
  \bibfield  {author} {\bibinfo {author} {\bibfnamefont {G.}~\bibnamefont
  {Scarcelli}}, \bibinfo {author} {\bibfnamefont {A.}~\bibnamefont {Valencia}},
  \bibinfo {author} {\bibfnamefont {S.}~\bibnamefont {Gompers}},\ and\ \bibinfo
  {author} {\bibfnamefont {Y.}~\bibnamefont {Shih}},\ }\bibfield  {title}
  {\bibinfo {title} {{Remote spectral measurement using entangled photons}},\
  }\href {https://doi.org/10.1063/1.1637131} {\bibfield  {journal} {\bibinfo
  {journal} {App. Phys. Lett.}\ }\textbf {\bibinfo {volume} {83}},\ \bibinfo
  {pages} {5560} (\bibinfo {year} {2003})},\ \Eprint
  {https://arxiv.org/abs/0407164} {arXiv:0407164 [quant-ph]} \BibitemShut
  {NoStop}%
\bibitem [{\citenamefont {Janassek}\ \emph {et~al.}(2018)\citenamefont
  {Janassek}, \citenamefont {Blumenstein},\ and\ \citenamefont
  {Els{\"{a}}{\ss}er}}]{Janassek2018}%
  \BibitemOpen
  \bibfield  {author} {\bibinfo {author} {\bibfnamefont {P.}~\bibnamefont
  {Janassek}}, \bibinfo {author} {\bibfnamefont {S.}~\bibnamefont
  {Blumenstein}},\ and\ \bibinfo {author} {\bibfnamefont {W.}~\bibnamefont
  {Els{\"{a}}{\ss}er}},\ }\bibfield  {title} {\bibinfo {title} {{Ghost
  Spectroscopy with Classical Thermal Light Emitted by a Superluminescent
  Diode}},\ }\bibfield  {journal} {\bibinfo  {journal} {Phys. Rev. Appl.}\
  }\textbf {\bibinfo {volume} {9}},\ \href
  {https://doi.org/10.1103/PhysRevApplied.9.021001}
  {10.1103/PhysRevApplied.9.021001} (\bibinfo {year} {2018})\BibitemShut
  {NoStop}%
\bibitem [{\citenamefont {Olivieri}\ \emph {et~al.}(2018)\citenamefont
  {Olivieri}, \citenamefont {{Totero Gongora}}, \citenamefont {Pasquazi},\ and\
  \citenamefont {Peccianti}}]{Olivieri2018}%
  \BibitemOpen
  \bibfield  {author} {\bibinfo {author} {\bibfnamefont {L.}~\bibnamefont
  {Olivieri}}, \bibinfo {author} {\bibfnamefont {J.~S.}\ \bibnamefont {{Totero
  Gongora}}}, \bibinfo {author} {\bibfnamefont {A.}~\bibnamefont {Pasquazi}},\
  and\ \bibinfo {author} {\bibfnamefont {M.}~\bibnamefont {Peccianti}},\
  }\bibfield  {title} {\bibinfo {title} {{Time-Resolved Nonlinear Ghost
  Imaging}},\ }\href {https://doi.org/10.1021/acsphotonics.8b00653} {\bibfield
  {journal} {\bibinfo  {journal} {ACS Photonics}\ }\textbf {\bibinfo {volume}
  {5}},\ \bibinfo {pages} {3379} (\bibinfo {year} {2018})}\BibitemShut
  {NoStop}%
\bibitem [{\citenamefont {Bromberg}\ \emph {et~al.}(2009)\citenamefont
  {Bromberg}, \citenamefont {Katz},\ and\ \citenamefont
  {Silberberg}}]{Bromberg2009}%
  \BibitemOpen
  \bibfield  {author} {\bibinfo {author} {\bibfnamefont {Y.}~\bibnamefont
  {Bromberg}}, \bibinfo {author} {\bibfnamefont {O.}~\bibnamefont {Katz}},\
  and\ \bibinfo {author} {\bibfnamefont {Y.}~\bibnamefont {Silberberg}},\
  }\bibfield  {title} {\bibinfo {title} {{Ghost imaging with a single
  detector}},\ }\href {https://doi.org/10.1103/PhysRevA.79.053840} {\bibfield
  {journal} {\bibinfo  {journal} {Phys. Rev. A}\ }\textbf {\bibinfo {volume}
  {79}},\ \bibinfo {pages} {053840} (\bibinfo {year} {2009})}\BibitemShut
  {NoStop}%
\bibitem [{\citenamefont {Clemente}\ \emph {et~al.}(2012)\citenamefont
  {Clemente}, \citenamefont {Dur\'an}, \citenamefont {Tajahuerce},
  \citenamefont {Torres-Company},\ and\ \citenamefont {Lancis}}]{Clemente2012}%
  \BibitemOpen
  \bibfield  {author} {\bibinfo {author} {\bibfnamefont {P.}~\bibnamefont
  {Clemente}}, \bibinfo {author} {\bibfnamefont {V.}~\bibnamefont {Dur\'an}},
  \bibinfo {author} {\bibfnamefont {E.}~\bibnamefont {Tajahuerce}}, \bibinfo
  {author} {\bibfnamefont {V.}~\bibnamefont {Torres-Company}},\ and\ \bibinfo
  {author} {\bibfnamefont {J.}~\bibnamefont {Lancis}},\ }\bibfield  {title}
  {\bibinfo {title} {Single-pixel digital ghost holography},\ }\href
  {https://doi.org/10.1103/PhysRevA.86.041803} {\bibfield  {journal} {\bibinfo
  {journal} {Phys. Rev. A}\ }\textbf {\bibinfo {volume} {86}},\ \bibinfo
  {pages} {041803} (\bibinfo {year} {2012})}\BibitemShut {NoStop}%
\bibitem [{\citenamefont {Howland}\ \emph {et~al.}(2011)\citenamefont
  {Howland}, \citenamefont {Dixon},\ and\ \citenamefont
  {Howell}}]{Howland2011}%
  \BibitemOpen
  \bibfield  {author} {\bibinfo {author} {\bibfnamefont {G.~A.}\ \bibnamefont
  {Howland}}, \bibinfo {author} {\bibfnamefont {P.~B.}\ \bibnamefont {Dixon}},\
  and\ \bibinfo {author} {\bibfnamefont {J.~C.}\ \bibnamefont {Howell}},\
  }\bibfield  {title} {\bibinfo {title} {{Photon-counting compressive sensing
  laser radar for 3D imaging}},\ }\href {https://doi.org/10.1364/AO.50.005917}
  {\bibfield  {journal} {\bibinfo  {journal} {Appl. Opt.}\ }\textbf {\bibinfo
  {volume} {50}},\ \bibinfo {pages} {5917} (\bibinfo {year}
  {2011})}\BibitemShut {NoStop}%
\bibitem [{\citenamefont {Howland}\ \emph {et~al.}(2013)\citenamefont
  {Howland}, \citenamefont {Lum}, \citenamefont {Ware},\ and\ \citenamefont
  {Howell}}]{Howland2013}%
  \BibitemOpen
  \bibfield  {author} {\bibinfo {author} {\bibfnamefont {G.~A.}\ \bibnamefont
  {Howland}}, \bibinfo {author} {\bibfnamefont {D.~J.}\ \bibnamefont {Lum}},
  \bibinfo {author} {\bibfnamefont {M.~R.}\ \bibnamefont {Ware}},\ and\
  \bibinfo {author} {\bibfnamefont {J.~C.}\ \bibnamefont {Howell}},\ }\bibfield
   {title} {\bibinfo {title} {{Photon counting compressive depth mapping}},\
  }\href {https://doi.org/10.1364/oe.21.023822} {\bibfield  {journal} {\bibinfo
   {journal} {Opt. Exp.}\ }\textbf {\bibinfo {volume} {21}},\ \bibinfo {pages}
  {23822} (\bibinfo {year} {2013})},\ \Eprint {https://arxiv.org/abs/1309.4385}
  {arXiv:1309.4385} \BibitemShut {NoStop}%
\bibitem [{\citenamefont {Chen}\ \emph {et~al.}(2013)\citenamefont {Chen},
  \citenamefont {Li}, \citenamefont {Gong}, \citenamefont {Bo}, \citenamefont
  {Xu}, \citenamefont {Zhao}, \citenamefont {Shen}, \citenamefont {Xu},\ and\
  \citenamefont {Han}}]{Chen2013}%
  \BibitemOpen
  \bibfield  {author} {\bibinfo {author} {\bibfnamefont {M.}~\bibnamefont
  {Chen}}, \bibinfo {author} {\bibfnamefont {E.}~\bibnamefont {Li}}, \bibinfo
  {author} {\bibfnamefont {W.}~\bibnamefont {Gong}}, \bibinfo {author}
  {\bibfnamefont {Z.}~\bibnamefont {Bo}}, \bibinfo {author} {\bibfnamefont
  {X.}~\bibnamefont {Xu}}, \bibinfo {author} {\bibfnamefont {C.}~\bibnamefont
  {Zhao}}, \bibinfo {author} {\bibfnamefont {X.}~\bibnamefont {Shen}}, \bibinfo
  {author} {\bibfnamefont {W.}~\bibnamefont {Xu}},\ and\ \bibinfo {author}
  {\bibfnamefont {S.}~\bibnamefont {Han}},\ }\bibfield  {title} {\bibinfo
  {title} {{Ghost Imaging Lidar via Sparsity Constraints in Real Atmosphere}},\
  }\href {https://doi.org/10.4236/opj.2013.32b021} {\bibfield  {journal}
  {\bibinfo  {journal} {Opt. Photonics J.}\ }\textbf {\bibinfo {volume} {03}},\
  \bibinfo {pages} {83} (\bibinfo {year} {2013})}\BibitemShut {NoStop}%
\bibitem [{\citenamefont {Yu}\ \emph {et~al.}(2015)\citenamefont {Yu},
  \citenamefont {Li}, \citenamefont {Gong},\ and\ \citenamefont
  {Han}}]{Yu2015}%
  \BibitemOpen
  \bibfield  {author} {\bibinfo {author} {\bibfnamefont {H.}~\bibnamefont
  {Yu}}, \bibinfo {author} {\bibfnamefont {E.}~\bibnamefont {Li}}, \bibinfo
  {author} {\bibfnamefont {W.}~\bibnamefont {Gong}},\ and\ \bibinfo {author}
  {\bibfnamefont {S.}~\bibnamefont {Han}},\ }\bibfield  {title} {\bibinfo
  {title} {{Structured image reconstruction for three-dimensional ghost imaging
  lidar}},\ }\href {https://doi.org/10.1364/oe.23.014541} {\bibfield  {journal}
  {\bibinfo  {journal} {Opt. Exp.}\ }\textbf {\bibinfo {volume} {23}},\
  \bibinfo {pages} {14541} (\bibinfo {year} {2015})}\BibitemShut {NoStop}%
\bibitem [{\citenamefont {Gong}\ \emph {et~al.}(2016)\citenamefont {Gong},
  \citenamefont {Zhao}, \citenamefont {Yu}, \citenamefont {Chen}, \citenamefont
  {Xu},\ and\ \citenamefont {Han}}]{Gong2016}%
  \BibitemOpen
  \bibfield  {author} {\bibinfo {author} {\bibfnamefont {W.}~\bibnamefont
  {Gong}}, \bibinfo {author} {\bibfnamefont {C.}~\bibnamefont {Zhao}}, \bibinfo
  {author} {\bibfnamefont {H.}~\bibnamefont {Yu}}, \bibinfo {author}
  {\bibfnamefont {M.}~\bibnamefont {Chen}}, \bibinfo {author} {\bibfnamefont
  {W.}~\bibnamefont {Xu}},\ and\ \bibinfo {author} {\bibfnamefont
  {S.}~\bibnamefont {Han}},\ }\bibfield  {title} {\bibinfo {title}
  {{Three-dimensional ghost imaging lidar via sparsity constraint}},\ }\href
  {https://doi.org/10.1038/srep26133} {\bibfield  {journal} {\bibinfo
  {journal} {Sci. Rep.}\ }\textbf {\bibinfo {volume} {6}},\ \bibinfo {pages}
  {1} (\bibinfo {year} {2016})}\BibitemShut {NoStop}%
\bibitem [{\citenamefont {Sun}\ \emph {et~al.}(2016)\citenamefont {Sun},
  \citenamefont {Edgar}, \citenamefont {Gibson}, \citenamefont {Sun},
  \citenamefont {Radwell}, \citenamefont {Lamb},\ and\ \citenamefont
  {Padgett}}]{Sun2016}%
  \BibitemOpen
  \bibfield  {author} {\bibinfo {author} {\bibfnamefont {M.~J.}\ \bibnamefont
  {Sun}}, \bibinfo {author} {\bibfnamefont {M.~P.}\ \bibnamefont {Edgar}},
  \bibinfo {author} {\bibfnamefont {G.~M.}\ \bibnamefont {Gibson}}, \bibinfo
  {author} {\bibfnamefont {B.}~\bibnamefont {Sun}}, \bibinfo {author}
  {\bibfnamefont {N.}~\bibnamefont {Radwell}}, \bibinfo {author} {\bibfnamefont
  {R.}~\bibnamefont {Lamb}},\ and\ \bibinfo {author} {\bibfnamefont {M.~J.}\
  \bibnamefont {Padgett}},\ }\bibfield  {title} {\bibinfo {title}
  {{Single-pixel three-dimensional imaging with time-based depth resolution}},\
  }\bibfield  {journal} {\bibinfo  {journal} {Nat. Commun.}\ }\textbf {\bibinfo
  {volume} {7}},\ \href {https://doi.org/10.1038/ncomms12010}
  {10.1038/ncomms12010} (\bibinfo {year} {2016}),\ \Eprint
  {https://arxiv.org/abs/1603.00726} {arXiv:1603.00726} \BibitemShut {NoStop}%
\bibitem [{\citenamefont {Musarra}\ \emph {et~al.}(2019)\citenamefont
  {Musarra}, \citenamefont {Lyons}, \citenamefont {Conca}, \citenamefont
  {Altmann}, \citenamefont {Villa}, \citenamefont {Zappa}, \citenamefont
  {Padgett},\ and\ \citenamefont {Faccio}}]{Musarra2019}%
  \BibitemOpen
  \bibfield  {author} {\bibinfo {author} {\bibfnamefont {G.}~\bibnamefont
  {Musarra}}, \bibinfo {author} {\bibfnamefont {A.}~\bibnamefont {Lyons}},
  \bibinfo {author} {\bibfnamefont {E.}~\bibnamefont {Conca}}, \bibinfo
  {author} {\bibfnamefont {Y.}~\bibnamefont {Altmann}}, \bibinfo {author}
  {\bibfnamefont {F.}~\bibnamefont {Villa}}, \bibinfo {author} {\bibfnamefont
  {F.}~\bibnamefont {Zappa}}, \bibinfo {author} {\bibfnamefont
  {M.}~\bibnamefont {Padgett}},\ and\ \bibinfo {author} {\bibfnamefont
  {D.}~\bibnamefont {Faccio}},\ }\bibfield  {title} {\bibinfo {title}
  {Non-line-of-sight three-dimensional imaging with a single-pixel camera},\
  }\href {https://doi.org/10.1103/PhysRevApplied.12.011002} {\bibfield
  {journal} {\bibinfo  {journal} {Phys. Rev. Applied}\ }\textbf {\bibinfo
  {volume} {12}},\ \bibinfo {pages} {011002} (\bibinfo {year}
  {2019})}\BibitemShut {NoStop}%
\bibitem [{\citenamefont {Aguilar}\ \emph {et~al.}(2019)\citenamefont
  {Aguilar}, \citenamefont {Hermosa},\ and\ \citenamefont
  {Soriano}}]{Aguilar2019}%
  \BibitemOpen
  \bibfield  {author} {\bibinfo {author} {\bibfnamefont {R.~A.}\ \bibnamefont
  {Aguilar}}, \bibinfo {author} {\bibfnamefont {N.}~\bibnamefont {Hermosa}},\
  and\ \bibinfo {author} {\bibfnamefont {M.~N.}\ \bibnamefont {Soriano}},\
  }\bibfield  {title} {\bibinfo {title} {{Low-cost Fourier ghost imaging using
  a light-dependent resistor}},\ }\href {https://doi.org/10.1119/10.0000163}
  {\bibfield  {journal} {\bibinfo  {journal} {Am. J. Phys.}\ }\textbf {\bibinfo
  {volume} {87}},\ \bibinfo {pages} {976} (\bibinfo {year} {2019})}\BibitemShut
  {NoStop}%
\bibitem [{\citenamefont {Sun}\ \emph {et~al.}(2013)\citenamefont {Sun},
  \citenamefont {Edgar}, \citenamefont {Bowman}, \citenamefont {Vittert},
  \citenamefont {Welsh}, \citenamefont {Bowman},\ and\ \citenamefont
  {Padgett}}]{Sun2013}%
  \BibitemOpen
  \bibfield  {author} {\bibinfo {author} {\bibfnamefont {B.}~\bibnamefont
  {Sun}}, \bibinfo {author} {\bibfnamefont {M.~P.}\ \bibnamefont {Edgar}},
  \bibinfo {author} {\bibfnamefont {R.}~\bibnamefont {Bowman}}, \bibinfo
  {author} {\bibfnamefont {L.~E.}\ \bibnamefont {Vittert}}, \bibinfo {author}
  {\bibfnamefont {S.}~\bibnamefont {Welsh}}, \bibinfo {author} {\bibfnamefont
  {A.}~\bibnamefont {Bowman}},\ and\ \bibinfo {author} {\bibfnamefont {M.~J.}\
  \bibnamefont {Padgett}},\ }\bibfield  {title} {\bibinfo {title} {{3D
  computational imaging with single-pixel detectors}},\ }\href
  {https://doi.org/10.1126/science.1234454} {\bibfield  {journal} {\bibinfo
  {journal} {Science}\ }\textbf {\bibinfo {volume} {340}},\ \bibinfo {pages}
  {844} (\bibinfo {year} {2013})}\BibitemShut {NoStop}%
\bibitem [{\citenamefont {Sun}\ and\ \citenamefont {Zhang}(2019)}]{Sun2019}%
  \BibitemOpen
  \bibfield  {author} {\bibinfo {author} {\bibfnamefont {M.-J.~J.}\
  \bibnamefont {Sun}}\ and\ \bibinfo {author} {\bibfnamefont {J.-M.~M.}\
  \bibnamefont {Zhang}},\ }\href {https://doi.org/10.3390/s19030732} {\bibinfo
  {title} {{Single-pixel imaging and its application in three-dimensional
  reconstruction: A brief review}}} (\bibinfo {year} {2019})\BibitemShut
  {NoStop}%
\bibitem [{\citenamefont {Zhang}\ and\ \citenamefont
  {Zhong}(2016)}]{Zhang2016}%
  \BibitemOpen
  \bibfield  {author} {\bibinfo {author} {\bibfnamefont {Z.}~\bibnamefont
  {Zhang}}\ and\ \bibinfo {author} {\bibfnamefont {J.}~\bibnamefont {Zhong}},\
  }\bibfield  {title} {\bibinfo {title} {{Three-dimensional single-pixel
  imaging with far fewer measurements than effective image pixels}},\ }\href
  {https://doi.org/10.1364/ol.41.002497} {\bibfield  {journal} {\bibinfo
  {journal} {Opt. Lett.}\ }\textbf {\bibinfo {volume} {41}},\ \bibinfo {pages}
  {2497} (\bibinfo {year} {2016})}\BibitemShut {NoStop}%
\bibitem [{\citenamefont {Zhang}\ \emph
  {et~al.}(2018{\natexlab{b}})\citenamefont {Zhang}, \citenamefont {Liu},
  \citenamefont {Peng}, \citenamefont {Yao}, \citenamefont {Zheng},\ and\
  \citenamefont {Zhong}}]{Zhang2018}%
  \BibitemOpen
  \bibfield  {author} {\bibinfo {author} {\bibfnamefont {Z.}~\bibnamefont
  {Zhang}}, \bibinfo {author} {\bibfnamefont {S.}~\bibnamefont {Liu}}, \bibinfo
  {author} {\bibfnamefont {J.}~\bibnamefont {Peng}}, \bibinfo {author}
  {\bibfnamefont {M.}~\bibnamefont {Yao}}, \bibinfo {author} {\bibfnamefont
  {G.}~\bibnamefont {Zheng}},\ and\ \bibinfo {author} {\bibfnamefont
  {J.}~\bibnamefont {Zhong}},\ }\bibfield  {title} {\bibinfo {title}
  {{Simultaneous spatial, spectral, and 3D compressive imaging via efficient
  Fourier single-pixel measurements}},\ }\href
  {https://doi.org/10.1364/optica.5.000315} {\bibfield  {journal} {\bibinfo
  {journal} {Optica}\ }\textbf {\bibinfo {volume} {5}},\ \bibinfo {pages} {315}
  (\bibinfo {year} {2018}{\natexlab{b}})}\BibitemShut {NoStop}%
\bibitem [{\citenamefont {Teng}\ \emph {et~al.}(2020)\citenamefont {Teng},
  \citenamefont {Guo}, \citenamefont {Chen}, \citenamefont {Yang},\ and\
  \citenamefont {Chen}}]{Teng2020}%
  \BibitemOpen
  \bibfield  {author} {\bibinfo {author} {\bibfnamefont {J.}~\bibnamefont
  {Teng}}, \bibinfo {author} {\bibfnamefont {Q.}~\bibnamefont {Guo}}, \bibinfo
  {author} {\bibfnamefont {M.}~\bibnamefont {Chen}}, \bibinfo {author}
  {\bibfnamefont {S.}~\bibnamefont {Yang}},\ and\ \bibinfo {author}
  {\bibfnamefont {H.}~\bibnamefont {Chen}},\ }\bibfield  {title} {\bibinfo
  {title} {{Time-encoded single-pixel 3D imaging}},\ }\href
  {https://doi.org/10.1063/1.5139924} {\bibfield  {journal} {\bibinfo
  {journal} {APL Photonics}\ }\textbf {\bibinfo {volume} {5}},\ \bibinfo
  {pages} {020801} (\bibinfo {year} {2020})},\ \Eprint
  {https://arxiv.org/abs/1901.04141} {arXiv:1901.04141} \BibitemShut {NoStop}%
\bibitem [{\citenamefont {Wang}\ \emph {et~al.}(2021)\citenamefont {Wang},
  \citenamefont {Bian},\ and\ \citenamefont {Zhang}}]{Wang2021}%
  \BibitemOpen
  \bibfield  {author} {\bibinfo {author} {\bibfnamefont {H.}~\bibnamefont
  {Wang}}, \bibinfo {author} {\bibfnamefont {L.}~\bibnamefont {Bian}},\ and\
  \bibinfo {author} {\bibfnamefont {J.}~\bibnamefont {Zhang}},\ }\bibfield
  {title} {\bibinfo {title} {Depth acquisition in single-pixel imaging with
  multiplexed illumination},\ }\href {https://doi.org/10.1364/OE.416481}
  {\bibfield  {journal} {\bibinfo  {journal} {Opt. Express}\ }\textbf {\bibinfo
  {volume} {29}},\ \bibinfo {pages} {4866} (\bibinfo {year}
  {2021})}\BibitemShut {NoStop}%
\bibitem [{\citenamefont {Qian}\ \emph {et~al.}(2019)\citenamefont {Qian},
  \citenamefont {He}, \citenamefont {Chen}, \citenamefont {Gu}, \citenamefont
  {Shi},\ and\ \citenamefont {Zhang}}]{Qian2019}%
  \BibitemOpen
  \bibfield  {author} {\bibinfo {author} {\bibfnamefont {Y.}~\bibnamefont
  {Qian}}, \bibinfo {author} {\bibfnamefont {R.}~\bibnamefont {He}}, \bibinfo
  {author} {\bibfnamefont {Q.}~\bibnamefont {Chen}}, \bibinfo {author}
  {\bibfnamefont {G.}~\bibnamefont {Gu}}, \bibinfo {author} {\bibfnamefont
  {F.}~\bibnamefont {Shi}},\ and\ \bibinfo {author} {\bibfnamefont
  {W.}~\bibnamefont {Zhang}},\ }\bibfield  {title} {\bibinfo {title} {{Adaptive
  compressed 3D ghost imaging based on the variation of surface normals}},\
  }\href {https://doi.org/10.1364/oe.27.027862} {\bibfield  {journal} {\bibinfo
   {journal} {Opt. Express}\ }\textbf {\bibinfo {volume} {27}},\ \bibinfo
  {pages} {27862} (\bibinfo {year} {2019})}\BibitemShut {NoStop}%
\bibitem [{\citenamefont {Zhang}\ \emph {et~al.}(2016)\citenamefont {Zhang},
  \citenamefont {Edgar}, \citenamefont {Sun}, \citenamefont {Radwell},
  \citenamefont {Gibson},\ and\ \citenamefont {Padgett}}]{Zhang2016a}%
  \BibitemOpen
  \bibfield  {author} {\bibinfo {author} {\bibfnamefont {Y.}~\bibnamefont
  {Zhang}}, \bibinfo {author} {\bibfnamefont {M.~P.}\ \bibnamefont {Edgar}},
  \bibinfo {author} {\bibfnamefont {B.}~\bibnamefont {Sun}}, \bibinfo {author}
  {\bibfnamefont {N.}~\bibnamefont {Radwell}}, \bibinfo {author} {\bibfnamefont
  {G.~M.}\ \bibnamefont {Gibson}},\ and\ \bibinfo {author} {\bibfnamefont
  {M.~J.}\ \bibnamefont {Padgett}},\ }\bibfield  {title} {\bibinfo {title} {{3D
  single-pixel video}},\ }\href {https://doi.org/10.1088/2040-8978/18/3/035203}
  {\bibfield  {journal} {\bibinfo  {journal} {J. Opt.}\ }\textbf {\bibinfo
  {volume} {18}} (\bibinfo {year} {2016})}\BibitemShut {NoStop}%
\bibitem [{\citenamefont {Soltanlou}\ and\ \citenamefont
  {Latifi}(2019)}]{Soltanlou2019}%
  \BibitemOpen
  \bibfield  {author} {\bibinfo {author} {\bibfnamefont {K.}~\bibnamefont
  {Soltanlou}}\ and\ \bibinfo {author} {\bibfnamefont {H.}~\bibnamefont
  {Latifi}},\ }\bibfield  {title} {\bibinfo {title} {{Three-dimensional imaging
  through scattering media using a single pixel detector}},\ }\href
  {https://doi.org/10.1364/ao.58.007716} {\bibfield  {journal} {\bibinfo
  {journal} {Appl. Opt.}\ }\textbf {\bibinfo {volume} {58}},\ \bibinfo {pages}
  {7716} (\bibinfo {year} {2019})}\BibitemShut {NoStop}%
\bibitem [{\citenamefont {Zhang}\ \emph {et~al.}(2019)\citenamefont {Zhang},
  \citenamefont {Lin}, \citenamefont {He}, \citenamefont {Qian}, \citenamefont
  {Chen},\ and\ \citenamefont {Zhang}}]{Zhang2019}%
  \BibitemOpen
  \bibfield  {author} {\bibinfo {author} {\bibfnamefont {L.}~\bibnamefont
  {Zhang}}, \bibinfo {author} {\bibfnamefont {Z.}~\bibnamefont {Lin}}, \bibinfo
  {author} {\bibfnamefont {R.}~\bibnamefont {He}}, \bibinfo {author}
  {\bibfnamefont {Y.}~\bibnamefont {Qian}}, \bibinfo {author} {\bibfnamefont
  {Q.}~\bibnamefont {Chen}},\ and\ \bibinfo {author} {\bibfnamefont
  {W.}~\bibnamefont {Zhang}},\ }\bibfield  {title} {\bibinfo {title}
  {{Improving the noise immunity of 3D computational ghost imaging}},\ }\href
  {https://doi.org/10.1364/oe.27.002344} {\bibfield  {journal} {\bibinfo
  {journal} {Opt. Express}\ }\textbf {\bibinfo {volume} {27}},\ \bibinfo
  {pages} {2344} (\bibinfo {year} {2019})}\BibitemShut {NoStop}%
\bibitem [{\citenamefont {Xi}\ \emph {et~al.}(2019)\citenamefont {Xi},
  \citenamefont {Chen}, \citenamefont {Yuan}, \citenamefont {Wang},
  \citenamefont {He}, \citenamefont {Liang}, \citenamefont {Liu}, \citenamefont
  {Zheng},\ and\ \citenamefont {Xu}}]{Xi2019}%
  \BibitemOpen
  \bibfield  {author} {\bibinfo {author} {\bibfnamefont {M.}~\bibnamefont
  {Xi}}, \bibinfo {author} {\bibfnamefont {H.}~\bibnamefont {Chen}}, \bibinfo
  {author} {\bibfnamefont {Y.}~\bibnamefont {Yuan}}, \bibinfo {author}
  {\bibfnamefont {G.}~\bibnamefont {Wang}}, \bibinfo {author} {\bibfnamefont
  {Y.}~\bibnamefont {He}}, \bibinfo {author} {\bibfnamefont {Y.}~\bibnamefont
  {Liang}}, \bibinfo {author} {\bibfnamefont {J.}~\bibnamefont {Liu}}, \bibinfo
  {author} {\bibfnamefont {H.}~\bibnamefont {Zheng}},\ and\ \bibinfo {author}
  {\bibfnamefont {Z.}~\bibnamefont {Xu}},\ }\bibfield  {title} {\bibinfo
  {title} {{Bi-frequency 3D ghost imaging with Haar wavelet transform}},\
  }\href {https://doi.org/10.1364/oe.27.032349} {\bibfield  {journal} {\bibinfo
   {journal} {Opt. Express}\ }\textbf {\bibinfo {volume} {27}},\ \bibinfo
  {pages} {32349} (\bibinfo {year} {2019})}\BibitemShut {NoStop}%
\bibitem [{\citenamefont {Woodham}(1980)}]{Woodham1980}%
  \BibitemOpen
  \bibfield  {author} {\bibinfo {author} {\bibfnamefont {R.~J.}\ \bibnamefont
  {Woodham}},\ }\bibfield  {title} {\bibinfo {title} {{Photometric Method For
  Determining Surface Orientation From Multiple Images}},\ }\href
  {https://doi.org/10.1117/12.7972479} {\bibfield  {journal} {\bibinfo
  {journal} {Opt. Eng.}\ }\textbf {\bibinfo {volume} {19}},\ \bibinfo {pages}
  {139} (\bibinfo {year} {1980})}\BibitemShut {NoStop}%
\bibitem [{\citenamefont {Horn}\ and\ \citenamefont {Brooks}(1989)}]{Horn1989}%
  \BibitemOpen
  \bibfield  {author} {\bibinfo {author} {\bibfnamefont {B.~K.~P.}\
  \bibnamefont {Horn}}\ and\ \bibinfo {author} {\bibfnamefont {M.~J.}\
  \bibnamefont {Brooks}},\ }\href@noop {} {\emph {\bibinfo {title} {{Shape From
  Shading}}}}\ (\bibinfo  {publisher} {MIT Press},\ \bibinfo {address}
  {Cambridge, CA, USA},\ \bibinfo {year} {1989})\ pp.\ \bibinfo {pages}
  {121--171}\BibitemShut {NoStop}%
\bibitem [{\citenamefont {Cho}\ \emph {et~al.}(2012)\citenamefont {Cho},
  \citenamefont {Matsushita}, \citenamefont {Tai},\ and\ \citenamefont
  {Kweon}}]{Cho2012}%
  \BibitemOpen
  \bibfield  {author} {\bibinfo {author} {\bibfnamefont {D.}~\bibnamefont
  {Cho}}, \bibinfo {author} {\bibfnamefont {Y.}~\bibnamefont {Matsushita}},
  \bibinfo {author} {\bibfnamefont {Y.~W.}\ \bibnamefont {Tai}},\ and\ \bibinfo
  {author} {\bibfnamefont {I.~S.}\ \bibnamefont {Kweon}},\ }\bibfield  {title}
  {\bibinfo {title} {{Semi-Calibrated Photometric Stereo}},\ }\href
  {https://doi.org/10.1109/TPAMI.2018.2873295} {\bibfield  {journal} {\bibinfo
  {journal} {IEEE Trans. Pattern Anal. Mach. Intell.}\ }\textbf {\bibinfo
  {volume} {42}},\ \bibinfo {pages} {232} (\bibinfo {year} {2012})}\BibitemShut
  {NoStop}%
\bibitem [{\citenamefont {Xiong}\ \emph {et~al.}(2015)\citenamefont {Xiong},
  \citenamefont {Chakrabarti}, \citenamefont {Basri}, \citenamefont {Gortler},
  \citenamefont {Jacobs},\ and\ \citenamefont {Zickler}}]{Xiong2015}%
  \BibitemOpen
  \bibfield  {author} {\bibinfo {author} {\bibfnamefont {Y.}~\bibnamefont
  {Xiong}}, \bibinfo {author} {\bibfnamefont {A.}~\bibnamefont {Chakrabarti}},
  \bibinfo {author} {\bibfnamefont {R.}~\bibnamefont {Basri}}, \bibinfo
  {author} {\bibfnamefont {S.~J.}\ \bibnamefont {Gortler}}, \bibinfo {author}
  {\bibfnamefont {D.~W.}\ \bibnamefont {Jacobs}},\ and\ \bibinfo {author}
  {\bibfnamefont {T.}~\bibnamefont {Zickler}},\ }\bibfield  {title} {\bibinfo
  {title} {{From shading to local shape}},\ }\href
  {https://doi.org/10.1109/TPAMI.2014.2343211} {\bibfield  {journal} {\bibinfo
  {journal} {IEEE Trans. Pattern Anal. Mach. Intell.}\ }\textbf {\bibinfo
  {volume} {37}},\ \bibinfo {pages} {67} (\bibinfo {year} {2015})},\ \Eprint
  {https://arxiv.org/abs/1310.2916} {arXiv:1310.2916} \BibitemShut {NoStop}%
\bibitem [{\citenamefont {Levenberg}(1944)}]{Levenberg1944}%
  \BibitemOpen
  \bibfield  {author} {\bibinfo {author} {\bibfnamefont {K.}~\bibnamefont
  {Levenberg}},\ }\bibfield  {title} {\bibinfo {title} {{A method for the
  solution of certain non-linear problems in least squares}},\ }\href@noop {}
  {\bibfield  {journal} {\bibinfo  {journal} {Q. Appl. Math.}\ }\textbf
  {\bibinfo {volume} {2}},\ \bibinfo {pages} {164} (\bibinfo {year}
  {1944})}\BibitemShut {NoStop}%
\bibitem [{\citenamefont {Marquardt}(1963)}]{Marquardt1963}%
  \BibitemOpen
  \bibfield  {author} {\bibinfo {author} {\bibfnamefont {D.~W.}\ \bibnamefont
  {Marquardt}},\ }\bibfield  {title} {\bibinfo {title} {{An Algorithm for
  Least-Squares Estimation of Nonlinear Parameters}},\ }\href
  {https://doi.org/10.1137/0111030} {\bibfield  {journal} {\bibinfo  {journal}
  {J. Soc. Ind. Appl. Math.}\ }\textbf {\bibinfo {volume} {11}},\ \bibinfo
  {pages} {431} (\bibinfo {year} {1963})}\BibitemShut {NoStop}%
\bibitem [{\citenamefont {Kanzow}\ \emph {et~al.}(2004)\citenamefont {Kanzow},
  \citenamefont {Yamashita},\ and\ \citenamefont {Fukushima}}]{Kanzow2004}%
  \BibitemOpen
  \bibfield  {author} {\bibinfo {author} {\bibfnamefont {C.}~\bibnamefont
  {Kanzow}}, \bibinfo {author} {\bibfnamefont {N.}~\bibnamefont {Yamashita}},\
  and\ \bibinfo {author} {\bibfnamefont {M.}~\bibnamefont {Fukushima}},\
  }\bibfield  {title} {\bibinfo {title} {{Levenberg-Marquardt methods with
  strong local convergence properties for solving nonlinear equations with
  convex constraints}},\ }\href {https://doi.org/10.1016/j.cam.2004.02.013}
  {\bibfield  {journal} {\bibinfo  {journal} {J. Comput. Appl. Math}\ }\textbf
  {\bibinfo {volume} {172}},\ \bibinfo {pages} {375} (\bibinfo {year}
  {2004})}\BibitemShut {NoStop}%
\bibitem [{\citenamefont {Frankot}\ and\ \citenamefont
  {Chellappa}(1988)}]{Frankot1988}%
  \BibitemOpen
  \bibfield  {author} {\bibinfo {author} {\bibfnamefont {R.~T.}\ \bibnamefont
  {Frankot}}\ and\ \bibinfo {author} {\bibfnamefont {R.}~\bibnamefont
  {Chellappa}},\ }\bibfield  {title} {\bibinfo {title} {{A Method for Enforcing
  Integrability in Shape from Shading Algorithms}},\ }\href
  {https://doi.org/10.1109/34.3909} {\bibfield  {journal} {\bibinfo  {journal}
  {IEEE Trans. Pattern Anal. Mach. Intell.}\ }\textbf {\bibinfo {volume}
  {10}},\ \bibinfo {pages} {439} (\bibinfo {year} {1988})}\BibitemShut
  {NoStop}%
\bibitem [{\citenamefont {Zhang}\ \emph {et~al.}(2015)\citenamefont {Zhang},
  \citenamefont {Ma},\ and\ \citenamefont {Zhong}}]{Zhang2015}%
  \BibitemOpen
  \bibfield  {author} {\bibinfo {author} {\bibfnamefont {Z.}~\bibnamefont
  {Zhang}}, \bibinfo {author} {\bibfnamefont {X.}~\bibnamefont {Ma}},\ and\
  \bibinfo {author} {\bibfnamefont {J.}~\bibnamefont {Zhong}},\ }\bibfield
  {title} {\bibinfo {title} {{Single-pixel imaging by means of Fourier spectrum
  acquisition}},\ }\href {https://doi.org/10.1038/ncomms7225} {\bibfield
  {journal} {\bibinfo  {journal} {Nat. Commun.}\ }\textbf {\bibinfo {volume}
  {6}},\ \bibinfo {pages} {6225} (\bibinfo {year} {2015})}\BibitemShut
  {NoStop}%
\bibitem [{\citenamefont {Taubman}\ and\ \citenamefont
  {Marcellin}(2002)}]{Taubman2002}%
  \BibitemOpen
  \bibfield  {author} {\bibinfo {author} {\bibfnamefont {D.~S.}\ \bibnamefont
  {Taubman}}\ and\ \bibinfo {author} {\bibfnamefont {M.~W.}\ \bibnamefont
  {Marcellin}},\ }\href@noop {} {\emph {\bibinfo {title} {{JPEG2000: Image
  Compression Fundamentals, Standards and Practice}}}}\ (\bibinfo  {publisher}
  {Springer US},\ \bibinfo {year} {2002})\BibitemShut {NoStop}%
\bibitem [{\citenamefont {Bian}\ \emph {et~al.}(2016)\citenamefont {Bian},
  \citenamefont {Suo}, \citenamefont {Hu}, \citenamefont {Chen},\ and\
  \citenamefont {Dai}}]{Bian2016}%
  \BibitemOpen
  \bibfield  {author} {\bibinfo {author} {\bibfnamefont {L.}~\bibnamefont
  {Bian}}, \bibinfo {author} {\bibfnamefont {J.}~\bibnamefont {Suo}}, \bibinfo
  {author} {\bibfnamefont {X.}~\bibnamefont {Hu}}, \bibinfo {author}
  {\bibfnamefont {F.}~\bibnamefont {Chen}},\ and\ \bibinfo {author}
  {\bibfnamefont {Q.}~\bibnamefont {Dai}},\ }\bibfield  {title} {\bibinfo
  {title} {{Efficient single pixel imaging in Fourier space}},\ }\href
  {https://doi.org/10.1088/2040-8978/18/8/085704} {\bibfield  {journal}
  {\bibinfo  {journal} {Journal of Optics}\ }\textbf {\bibinfo {volume} {18}},\
  \bibinfo {pages} {085704} (\bibinfo {year} {2016})}\BibitemShut {NoStop}%
\bibitem [{\citenamefont {Slepian}(1965)}]{Slepian1965}%
  \BibitemOpen
  \bibfield  {author} {\bibinfo {author} {\bibfnamefont {D.}~\bibnamefont
  {Slepian}},\ }\bibfield  {title} {\bibinfo {title} {{Analytic Solution of Two
  Apodization Problems}},\ }\href {https://doi.org/10.1364/josa.55.001110}
  {\bibfield  {journal} {\bibinfo  {journal} {JOSA}\ }\textbf {\bibinfo
  {volume} {55}},\ \bibinfo {pages} {1110} (\bibinfo {year}
  {1965})}\BibitemShut {NoStop}%
\bibitem [{\citenamefont {McCutchen}(1969)}]{McCutchen1969}%
  \BibitemOpen
  \bibfield  {author} {\bibinfo {author} {\bibfnamefont {C.~W.}\ \bibnamefont
  {McCutchen}},\ }\bibfield  {title} {\bibinfo {title} {{Two Families of
  Apodization Problems}},\ }\href {https://doi.org/10.1364/JOSA.59.001163}
  {\bibfield  {journal} {\bibinfo  {journal} {JOSA}\ }\textbf {\bibinfo
  {volume} {59}},\ \bibinfo {pages} {1163} (\bibinfo {year}
  {1969})}\BibitemShut {NoStop}%
\bibitem [{\citenamefont {Thomas}\ and\ \citenamefont
  {Gadhok}(2001)}]{Thomas2001}%
  \BibitemOpen
  \bibfield  {author} {\bibinfo {author} {\bibfnamefont {G.}~\bibnamefont
  {Thomas}}\ and\ \bibinfo {author} {\bibfnamefont {N.}~\bibnamefont
  {Gadhok}},\ }\bibfield  {title} {\bibinfo {title} {{Sidelobe apodization in
  Fourier imaging}},\ }in\ \href {https://doi.org/10.1109/acssc.2001.987715}
  {\emph {\bibinfo {booktitle} {Conf. Rec. Thirty-Fifth Asilomar Conf. Signals,
  Syst. and Comput.}}},\ Vol.~\bibinfo {volume} {2}\ (\bibinfo {year} {2001})\
  pp.\ \bibinfo {pages} {1369--1373}\BibitemShut {NoStop}%
\end{thebibliography}%
	
\end{document}